# REVIEW OF THE CLASSIFICATION AND PROPERTIES OF 40 VARIABLE STARS IN SAGITTARIUS

AGLI' GIOELE[1], ARCOMANO GAIA[1], BELLO GABRIELE[1], BOCCHINO VALENTINO[1],
BRUNO GAIA[1], CARRO CHRISTIAN[1], CHARRIER MATTEO[1], FANTINO FILIPPO[1],
HU GIOVANNI[1], RICCO MARTINA[1], AND UGHETTO-MONFRIN MATTIA[1],
BENNA CARLO[2], GARDIOL DANIELE[2] AND PETTITI GIUSEPPE[2]

1) Liceo Scientifico Maria Curie, Via dei Rochis 12, I-10064 Pinerolo (TO), Italy, TOPS070007@istruzione.it

2) INAF-Osservatorio Astrofisico di Torino, via Osservatorio 20, I-10025 Pino Torinese (TO), Italy, giuseppe.pettiti@inaf.it

**Abstract**. This study aims to assess the properties and classification of 40 variable stars in Sagittarius, little studied since their discovery and reported in the Information Bulletin on Variable Stars (IBVS) 985 and update.

Using data from previous studies and several astronomical databases, we performed our analysis mainly utilizing a period analysis software and comparing the photometric characteristics of the variables in a Colour-Absolute Magnitude Diagram.

For all stars, the variability is confirmed. We discovered new significant results for the period and/or type of 15 variables and highlighted incorrect cross-reference names on astronomical databases for 5 stars. This assessment also identifies 9 cases for which results from the ASAS-SN Catalog of Variable Stars are systematically not consistent with the original light curves. A correct identification of NSV 10522 is provided.

## 1. Introduction

The IBVS 985 (Maffei, 1975) describes the results of a systematic photographic search of variable stars in a field centered at R.A. = $18^h\ 14^m$, Dec. = $-14°\ 50'$ (1900.0), containing the objects M16 and M17, performed in blue and infrared band at the Astrophysical Observatory of Asiago beginning from the summer of 1967. This photographic survey led to the discovery of 197 new variables, mostly late spectral type stars with low effective temperature and maximum of luminosity at infrared wavelengths, confirming the strong increase of the number of Mira type stars when the observations are made in the infrared band.

An update of the characteristics of 176 Mira and Semiregular variable stars originally observed, an in particular their revised coordinates and finding charts, were provided with a later paper (Maffei and Tosti, 2013). The update made use of discontinuous observations performed with the two Schmidt telescopes of the Astrophysical Observatory of Asiago and with the Schmidt telescope of the Catania Astrophysical Observatory from 1961 to 1991, with photographic emulsion 103a-O without or with filter GG13 and hypersensitized with $NH_3$ or pre-flashed I-N+RG5. A revision of variables identification which provides 2MASS counterparts for 19 stars was issued in 2018 (Nesci, 2018).

Initial classification and properties of variable stars discovered through surveys are sometimes affected by mistakes or uncertainties that remain undiscovered or unresolved due to the lack of new observations or further analysis. The variable stars of IBVS 985 are distributed in the constellations of Sagittarius, Serpens and Scutum and are generally faint and little studied since their discovery.

In this study, we assess the original physical and photometric characteristics and the classification of 40 variable stars of IBVS 985 belonging to the constellation of Sagittarius, based on data mined by more recent and public astronomical articles and databases. The provisional name, original coordinates, final designation, and main characteristics of the stars covered by this study are shown in Table 1.

In this paper, when the term 'original' is used for the characteristics of the variables, we refer to the updated data defined in the Atlas of 176 Mira/SR Light Curves (Maffei and Tosti, 2013).

Section 2 describes the method of our assessment. In section 2.1, we provide a general assessment of the type of the variables based on their Gaia photometric characteristics; a more detailed analysis for each variable, based on the results of previous studies and of our photometric analysis, is provided in section 2.2.

Our conclusions are summarized in section 3.

Table 1    Original main characteristics of the variable stars in Sagittarius

| Provisional name | Original coordinates R.A. - Decl. (1950.0) | Variable Designation | Type | Mag. Range (I-N + RG5) | Period (d) |
|---|---|---|---|---|---|
| M016 | 18 10 41   -16 14.7 | V3908 Sgr | M | 8.4 ÷ 11.1 | 274.5 |
| M023 | 18 24 30   -16 17.5 | V3945 Sgr | M | 14.1 ÷ 17.1 | 321.0 |
| M029 | 18 14 40   -17 43.4 | V3921 Sgr | M | 13.3 ÷ 15.9 | 531.0 |
| M035 | 18 15 35   -17 26.9 | NSV 10522 | --- | 15.5 ÷ 16.8 | --- |
| M036 | 18 15 54   -17 45.1 | V3923 Sgr | M:: | 13.3 ÷ <15.6 | --- |
| M037 | 18 16 33   -17 49.1 | V3925 Sgr | M | 12.1 ÷ 15.8 | 151.5 |
| M047 | 18 18 48   -17 33.5 | V3932 Sgr | M | 12.3 ÷ 15.0 | 205.0 |
| M048 | 18 18 28   -16 44.4 | V3931 Sgr | M | 12.0 ÷ 15.6 | 441.0 |
| M052 | 18 21 15   -17 05.3 | V3935 Sgr | M | 14.1 ÷ 16.0 | 395.0 |
| M053 | 18 21 31   -17 05.6 | NSV 10741 | --- | 13.8 ÷ <16.3 | --- |
| M054 | 18 21 45   -17 06.7 | V3938 Sgr | M | 12.8 ÷ 15.8 | 297.0 |
| M057 | 18 22 16   -16 38.1 | V3940 Sgr | M | 11.6 ÷ 15.2 | 449.0 |
| M058 | 18 23 55   -16 23.6 | V3944 Sgr | M | 14.7 ÷ 17.2 | 286.0 |
| M059 | 18 24 48   -16 23.8 | V3946 Sgr | M | 13.8 ÷ 17.3 | 411.0 |
| M060 | 18 24 54   -16 09.0 | V3947 Sgr | M | 11.2 ÷ 15.4 | 314.5 |
| M086 | 18 13 52   -16 53.3 | V3918 Sgr | M | 12.5 ÷ 15.6 | 384.0 |
| M088 | 18 14 39   -17 33.2 | V3920 Sgr | SR:: | 14.3 ÷ 16.0 | --- |
| M089 | 18 08 13   -16 13.7 | V3904 Sgr | M | 12.6 ÷ 15.2 | 360.0 |
| M096 | 18 16 33   -17 47.1 | V3924 Sgr | M | 13.8 ÷ 16.0 | 372.0 |
| M097 | 18 17 02   -17 44.5 | V3927 Sgr | SR | 12.5 ÷ 15.4 | 218.0 |
| M108 | 18 18 14   -17 07.8 | V3930 Sgr | M | 13.7 ÷ 16.3 | 448.0 |
| M109 | 18 20 12   -17 22.7 | V3934 Sgr | SRa | 13.3 ÷ 14.8 | 366.0 |
| M111 | 18 19 06   -16 35.6 | NSV 10693 | --- | 14.8 ÷ 16.3 | --- |
| M112 | 18 21 24   -17 09.0 | V3936 Sgr | M | 13.3 ÷ 15.8 | 369.0 |
| M113 | 18 22 11   -17 05.7 | NSV 10752 | M | 14.6 ÷ 16.5 | 206.0 |

| Provisional name | Original coordinates R.A. - Decl. (1950.0) | Variable Designation | Type | Mag. Range (I-N + RG5) | Period (d) |
|---|---|---|---|---|---|
| M119 | 18 21 34  -16 11.8 | V3937 Sgr | M | 13.8 ÷ <17.2 | 606.0 |
| M120 | 18 22 46  -16 35.5 | V3942 Sgr | M | 11.2 ÷ 15.3 | 370.0 |
| M123 | 18 23 54  -15 59.4 | V3943 Sgr | M | 14.7 ÷ 17.4 | 374.0 |
| M124 | 18 24 59  -16 24.0 | V3948 Sgr | M | 14.4 ÷ <17.4 | 420.0 |
| M125 | 18 25 02  -16 23.5 | NSV 10837 | SRa | 16.1 ÷ <17.4 | 416.0 |
| M126 | 18 25 26  -16 23.1 | V3949 Sgr | M: | 13.2 ÷ 17.1 | 520.0 |
| M127 | 18 25 45  -16 05.4 | V3950 Sgr | SRa | 14.8 ÷ 16.9 | 198.0: |
| M161 | 18 14 24  -17 35.3 | NSV 10490 | E:: | 15.3 ÷ <15.9 | --- |
| M162 | 18 08 20  -16 04.1 | V3905 Sgr | SRa | 14.6 ÷ 16.5 | 328.0 |
| M168 | 18 16 39  -17 49.9 | V3926 Sgr | M | 12.4 ÷ 15.7 | 290.0 |
| M169 | 18 17 04  -17 32.9 | NSV 10626 | M: | 14.7 ÷ <16.0 | 338.0:: |
| M170 | 18 15 59  -17 19.8 | NSV 10577 | M | 14.0 ÷ <16.0 | 665.0 |
| M183 | 18 22 54  -16 13.5 | NSV 10772 | --- | 16.0 ÷ 16.8 | --- |
| M184 | 18 22 17  -16 12.3 | NSV 10757 | --- | 16.2 ÷ <17.2 | --- |
| M185 | 18 21 49  -16 10.8 | V3939 Sgr | M: | 15.5 ÷ <17.1 | 292.0 |

## 2. Data analysis

For each variable star listed in Table 1, we first checked the Gaia source identifier reported by the SIMBAD database or, when this information was missing, we identified a Gaia counterpart using the variable coordinates derived from the original finding charts. If more than one Gaia identifier was possible, due to the uncertainty on the variable known position, we selected the Gaia source based on its expected photometric characteristics. A rationale of the selection criterion for variable stars for which such approach was necessary is provided in the detailed data analysis of section 2.2.

Once the identity of all variable stars has been confirmed, in addition to the information available on SIMBAD, we looked for physical, spectroscopic, and photometric data available in other astronomical databases and catalogues, specifically the ones from the American Association of Variable Stars Observers (AAVSO, Kafka 2020), the All-Sky Automated Survey for Supernovae (ASAS-SN, Kochanek et al. 2017; Jayasinghe et al. 2019), the 2MASS All Sky Catalogue (Cutri et al. 2003), the Gaia Data Release 2 (Gaia Collaboration et al. 2016; Gaia Collaboration et al. 2018b), and the Gaia Early Data Release 3 (Gaia Collaboration et al. 2020a).

In Table 2 we list, for each variable, the Gaia DR2/EDR3 source ID and the apparent median Gaia G magnitude in DR2 and EDR3 with associated errors, because the photometric system for G, Bp and Rp magnitudes in Gaia EDR3 is different from the one used in Gaia DR2.

Table 3 shows the absolute magnitudes calculated using the distances available from the Gaia Distances to 1.33 billion stars in Gaia DR2 catalogue (Bailer-Jones et al. 2018) and also summarizes the effective temperature $T_{eff}$ and the Gaia colour index Bp-Rp. The interstellar extinction or circumstellar extinction was not considered in the absolute magnitude calculation because the extinction value estimation in the filter G is available only for 8 of our sample of 40 variable stars.

Table 2      Gaia DR2/EDR3 source ID and apparent median G magnitudes

| Name | Gaia DR2/EDR3 Source ID | Gaia DR2 Median G (mag.) | ΔG (±) | Gaia EDR3 Median G (mag.) | ΔG (±) |
|---|---|---|---|---|---|
| NSV 10490 | 4097207867633543808 | 15.149 | 0.033 | 15.095 | 0.025 |
| NSV 10522 | 4097214842660615808 | 16.573 | 0.101 | 16.805 | 0.096 |
| NSV 10577 | 4097310117901098880 | 15.165 | 0.104 | 16.478 | 0.150 |
| NSV 10626 | 4096540567217210368 | 15.866 | 0.056 | 15.772 | 0.057 |
| NSV 10693 | 4097737660433986816 | 15.809 | 0.057 | 15.689 | 0.040 |
| NSV 10741 | 4096948550376111616 | 13.531 | 0.001 | 13.503 | 0.003 |
| NSV 10752 | 4096900966474840320 | 15.704 | 0.047 | 15.509 | 0.038 |
| NSV 10757 | 4097858602406503552[1] <br> 4097858606748077696[2] | 15.755 | 0.023 | 17.134 | 0.090 |
| NSV 10772 | 4097856270277660032 | 14.817 | 0.005 | 14.804 | 0.004 |
| NSV 10837 | 4097091010218371840[1] <br> 4097091005854942848[2] | 14.799 | 0.019 | 16.106 | 0.093 |
| V3904 Sgr | 4145671007298104960 | 14.622 | 0.049 | 14.727 | 0.050 |
| V3905 Sgr | 4145677569997394688 | 16.321 | 0.063 | 16.094 | 0.060 |
| V3908 Sgr | 4145615030968253952 | 10.524 | 0.040 | 10.764 | 0.045 |
| V3918 Sgr | 4104382730793346688 | 13.966 | 0.039 | 13.941 | 0.033 |
| V3920 Sgr | 4097209551260714880 | 14.877 | 0.078 | 15.053 | 0.062 |
| V3921 Sgr | 4097202748032445312 | 15.082 | 0.103 | 15.150 | 0.062 |
| V3923 Sgr | 4097188106476834816 | 15.987 | 0.091 | 16.124 | 0.094 |
| V3924 Sgr | 4096442298393113856 | 15.944 | 0.073 | 15.642 | 0.053 |
| V3925 Sgr | 4096439274736278272 | 14.617 | 0.068 | 14.324 | 0.053 |
| V3926 Sgr | 4096438617497438720 | 14.374 | 0.050 | 14.190 | 0.042 |
| V3927 Sgr | 4096441198881091328 | 13.783 | 0.043 | 14.224 | 0.052 |
| V3930 Sgr | 4097329929300561280 | 16.605 | 0.031 | 15.842 | 0.077 |
| V3931 Sgr | 4097359462791432704 | 13.916 | 0.047 | 14.272 | 0.047 |
| V3932 Sgr | 4096532213423534336 | 14.163 | 0.038 | 14.218 | 0.032 |
| V3934 Sgr | 4096561664098938368 | 14.494 | 0.000 | 14.468 | 0.003 |
| V3935 Sgr | 4096942885361078912 | 15.850 | 0.045 | 15.789 | 0.045 |
| V3936 Sgr | 4096941751489519616 | 15.004 | 0.026 | 14.947 | 0.022 |
| V3937 Sgr | 4097777444701491584 | 15.732 | 0.076 | 15.793 | 0.062 |
| V3938 Sgr | 4096947592644959232 | 14.475 | 0.060 | 14.641 | 0.059 |
| V3939 Sgr | 4097871182403713152 | 15.519 | 0.068 | 15.747 | 0.055 |
| V3940 Sgr | 4097000571017265408 | 13.371 | 0.035 | 13.612 | 0.034 |
| V3942 Sgr | 4097002430790019712[1] <br> 4097002430790019840[2] | 12.261 | 0.010 | 14.172 | 0.044 |
| V3943 Sgr | 4097889461785848576 | 15.358 | 0.066 | 15.389 | 0.061 |
| V3944 Sgr | 4097087883482681344 | 15.553 | 0.056 | 15.769 | 0.054 |

| Name | Gaia DR2/EDR3 Source ID | Gaia DR2 Median G (mag.) | ΔG (±) | Gaia EDR3 Median G (mag.) | ΔG (±) |
|---|---|---|---|---|---|
| V3945 Sgr | 4097096056729738624 | 14.498 | 0.092 | 14.233 | 0.058 |
| V3946 Sgr | 4097090769700294656 | 15.355 | 0.071 | 15.238 | 0.061 |
| V3947 Sgr | 4097121070619720704 | 13.676 | 0.061 | 13.745 | 0.040 |
| V3948 Sgr | 4097090632223485824 | 15.706 | 0.093 | 15.434 | 0.064 |
| V3949 Sgr | 4097067336305154304 | 13.856 | 0.040 | 13.771 | 0.031 |
| V3950 Sgr | 4097128324897371264 | 14.570 | 0.016 | 14.616 | 0.012 |

1  Gaia DR2 source ID
2  Gaia EDR3 source ID

Table 3    Absolute Magnitude, Effective Temperature and color index

| Name | DR2 Absolute Magnitude ($M_G$) | Δ $M_G$ (±) | Teff (K) | Δ Teff (± K) | DR2 Bp-Rp |
|---|---|---|---|---|---|
| NSV 10490 | 5.77 | -1.29 ÷ 0.80 | 3284 | -8 ÷ 51 | 6.598 |
| NSV 10522 | 4.58 | -1.52 ÷ 1.31 | 3286 | -9 ÷ 91 | 5.190 |
| NSV 10577 | --- | --- | 3280 | -4 ÷ 8 | 7.110 |
| NSV 10626 | 2.88 | -1.15 ÷ 1.14 | --- | --- | --- |
| NSV 10693 | 3.74 | -1.51 ÷ 1.32 | 3284 | -8 ÷ 55 | 6.529 |
| NSV 10741 | 1.35 | -0.77 ÷ 0.60 | 4100 | -491 ÷ 670 | 3.107 |
| NSV 10752 | 1.80 | -0.89 ÷ 0.89 | 3295 | -16 ÷ 154 | 4.672 |
| NSV 10757 | --- | --- | 3284 | -9 ÷ 51 | 6.490 |
| NSV 10772 | 1.30 | -1.01 ÷ 1.03 | 3281 | -11 ÷ 29 | 6.354 |
| NSV 10837 | --- | --- | 3363 | -71 ÷ 346 | 4.156 |
| V3904 Sgr | 3.06 | -1.63 ÷ 1.19 | 3281 | -5 ÷ 8 | 7.060 |
| V3905 Sgr | 3.74 | -1.29 ÷ 1.31 | 3319 | -40 ÷ 78 | 5.288 |
| V3908 Sgr | 0.75 | -0.58 ÷ 0.46 | 3483 | -195 ÷ 872 | 5.922 |
| V3918 Sgr | 1.60 | -1.42 ÷ 1.32 | 3283 | -11 ÷ 41 | 6.637 |
| V3920 Sgr | 4.75 | -0.76 ÷ 0.57 | 3281 | -5 ÷ 12 | 6.335 |
| V3921 Sgr | 5.67 | -1.29 ÷ 0.81 | 3281 | -5 ÷ 7 | 6.516 |
| V3923 Sgr | 3.00 | -1.15 ÷ 1.13 | 3325 | -46 ÷ 75 | 5.284 |
| V3924 Sgr | 2.74 | -1.05 ÷ 1.01 | 3281 | -6 ÷ 10 | 5.919 |
| V3925 Sgr | 2.82 | -1.25 ÷ 0.90 | 3281 | -6 ÷ 9 | 5.995 |
| V3926 Sgr | 3.49 | -1.20 ÷ 0.80 | 3282 | -10 ÷ 53 | 5.981 |
| V3927 Sgr | 0.84 | -1.11 ÷1.00 | 3281 | -13 ÷ 38 | 5.719 |
| V3930 Sgr | 4.36 | -1.49 ÷ 1.58 | 3284 | -8 ÷ 33 | 5.642 |
| V3931 Sgr | 0.98 | -1.17 ÷ 1.16 | 3281 | -5 ÷ 8 | 7.212 |
| V3932 Sgr | 0.15 | -0.86 ÷ 0.87 | 3281 | -13 ÷ 38 | 5.854 |
| V3934 Sgr | 2.37 | -0.37 ÷ 0.32 | 3832 | -532 ÷ 743 | 2.565 |
| V3935 Sgr | 2.56 | -1.07 ÷ 1.07 | 3288 | -9 ÷ 71 | 5.560 |

| Name | DR2 Absolute Magnitude (M$_G$) | Δ M$_G$ (±) | Teff (K) | Δ Teff (± K) | DR2 Bp-Rp |
|---|---|---|---|---|---|
| V3936 Sgr | 2.44 | -1.33 ÷ 1.23 | 3284 | -9 ÷ 51 | 6.276 |
| V3937 Sgr | 5.49 | -2.07 ÷ 1.11 | 3281 | -5 ÷ 8 | 7.027 |
| V3938 Sgr | 1.70 | -1.21 ÷ 1.11 | 3283 | -9 ÷ 52 | 6.814 |
| V3939 Sgr | 5.94 | -0.69 ÷ 0.53 | 3283 | -8 ÷ 49 | 6.350 |
| V3940 Sgr | 2.79 | -2.26 ÷ 1.27 | 3281 | -5 ÷ 8 | 7.041 |
| V3942 Sgr | --- | --- | 3288 | -16 ÷ 36 | 5.827 |
| V3943 Sgr | 2.93 | -1.36 ÷ 1.19 | 3284 | -9 ÷ 51 | 6.502 |
| V3944 Sgr | 2.01 | -0.99 ÷ 0.99 | 3286 | -14 ÷ 53 | 5.591 |
| V3945 Sgr | 0.57 | -0.91 ÷ 0.93 | 3284 | -9 ÷ 35 | 5.737 |
| V3946 Sgr | 2.25 | -1.16 ÷ 1.16 | 3285 | -10 ÷ 57 | 5.747 |
| V3947 Sgr | 0.12 | -1.01 ÷ 1.03 | 3281 | -6 ÷ 6 | 6.090 |
| V3948 Sgr | 2.73 | -1.16 ÷ 1.08 | 3281 | -6 ÷ 56 | 5.036 |
| V3949 Sgr | 0.32 | -1.04 ÷ -1.08 | 3280 | -4 ÷ 8 | 6.614 |
| V3950 Sgr | 1.14 | -1.03 ÷ 1.00 | 3336 | -48 ÷ 79 | 4.887 |

Based on their Gaia EDR3 equatorial coordinates (transformed at J2000.0), we searched for each variable the photometric measurements and light curves analysis available from ASAS-SN database. If no variable is reported by ASAS-SN database we performed a period analysis of the photometric data of the source that matches the Gaia EDR3 equatorial coordinates. We analyzed the valid photometric data using version 2.51 and 2.60 of the light curve and period analysis software PERANSO (Paunzen and Vanmunster, 2016). The analysis of the period was performed using two methods: Lomb-Scargle (Lomb 1976, Scargle 1982) and ANOVA (Schwarzenberg-Czerny A., 1996), that are deemed powerful in detecting weak periodic signals, improving sensitivity of peak detection, and damping alias periods. For each value of the period, reported in this paper, a Fisher Randomization Test, with 200 permutations, was run with PERANSO software, in order to confirm the significance of the period we found. All results reported in this paper have a False Alarm Probabilities (FAP) less than 1%, indicating a high significance of the result. Any period with a FAP greater than 1% was disregarded.

During our search for photometric data, we found for 5 variables (see Table 4) a mismatch on the cross-reference of the Gaia EDR3 source ID reported by the ASAS-SN photometry database. We investigated the reason for this difference and more details of our analysis are provided for the single stars in section 2.2.

It is also noted that for some of the variables (see Table 5) the AAVSO database refers to IBVS 985 (Maffei 1975) data and should be updated to reflect the content of the later paper (Maffei and Tosti, 2013).

In general, for each star, bibliographic references available from SIMBAD were reviewed.

Table 4	Incorrect ASAS-SN Gaia EDR3 source ID

| Name | Gaia EDR3 Source ID | ASAS-SN Gaia EDR3 Source ID |
|---|---|---|
| NSV 10522 | 4097214842660615808 | 4097214842648260096 |
| NSV 10741 | 4096948550376111616 | 4096948550378843264 |
| V3920 Sgr | 4097209551260714880 | 4097209551260717312 |
| V3934 Sgr | 4096561664098938368 | 4096561659747632000 |
| V3947 Sgr | 4097121070619720704 | 4097121074968426752 |

Table 5	Variable periods in the AAVSO database vs later paper

|  | AAVSO (Maffei 1975) | Revised values (Maffei and Tosti, 2013) |
|---|---|---|
| Name | Period (d) | Period (d) |
| V3908 Sgr | 279.5 | 274.5 |
| V3923 Sgr | 396: | --- |
| V3926 Sgr | 286: | 290.0 |
| V3932 Sgr | 208 | 205.0 |
| V3937 Sgr | 608.0 | 606.0 |
| V3939 Sgr | 268 | 292.0 |
| V3942 Sgr | 372.0 | 370.0 |
| V3946 Sgr | 405 | 411.0 |
| V3947 Sgr | 312 | 314.5 |
| V3948 Sgr | 410.0 | 420.0 |
| V3949 Sgr | 510.0 | 520.0 |

## 2.1. Variable stars in the Gaia Colour-Absolute Magnitude Diagram

To get a first general indication of the type of the variable stars listed in Table 1, we first analyzed their position in a Gaia DR2 Colour-Absolute Magnitude Diagram (CAMD). Figure 1 shows the absolute magnitude $M_G$ vs. color index Bp-Rp for a training sample of about 166,000 variable stars. Different classes of variables are shown in different colors, each one occupying specific areas of the diagram (Jayasinghe et al. 2019). In the same figure, we reported the position of the stars of IBVS 985, as diamond-shaped dots. All photometric data reported in the diagram are not corrected for the interstellar and circumstellar extinction.

Figure 1 highlights that most of the variable stars of IBVS 985 have photometric characteristics compatible with those of a Long Period Variable (LPV) of the training sample i.e., belonging to the Mira, Semiregular or Irregular variable types.

In the same Figure, we have indicated with red squares the variables that are classified by WISE (Marton et al. 2016) as candidate Young Stellar Objects (YSO).

Due to the overlap of the various groups of variables in the CAMD, this comparison does not allow a fine classification of the class of variables, but it is still a valid indicator of whether a variable is a candidate to be a Long Period Variable or a YSO.

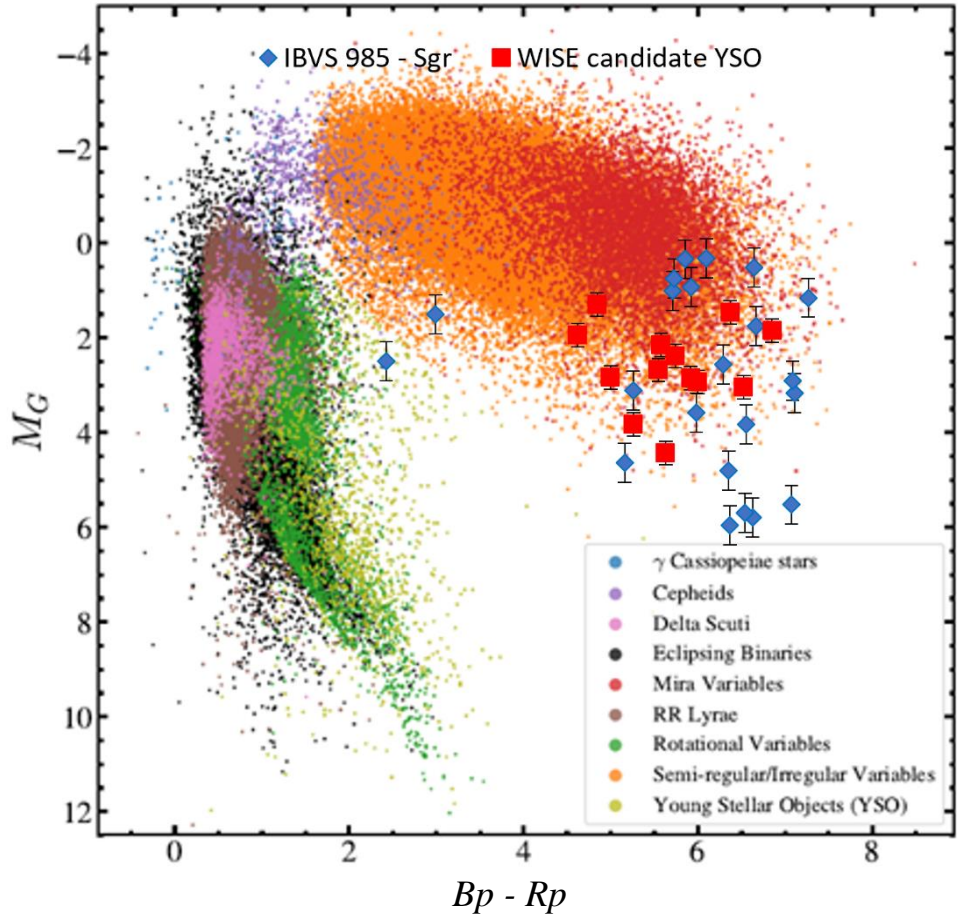

Figure 1    IBVS 985 variable stars in Sagittarius vs. Gaia DR2 CAMD

Based on their position in the Gaia DR2 CAMD, we have classified 35 out of 40 variables into the following groups, which are summarized in Table 6:
  a. Not-LPV stars. This group contains only the variable V3934 Sgr, whose color index Bp-Rp = 2.56 and absolute magnitude $M_G$ = 2.37 are not compatible with a LPV star.
  b. SR/IR. This group contains only the variable NSV 10741, whose photometric data are consistent with a Semiregular/Irregular variable.
  c. Probable YSO. This group contains seven stars with a value of absolute magnitude $M_G$ > 4.3, that are considered not compatible with a LPV star.
  d. LPV. These 26 stars show photometric characteristics compatible with a Long Period Variable.

For 5 of the 40 stars either the colour index Bp-Rp or the absolute magnitude $M_G$ is not available and therefore they cannot be classified with respect to the CAMD.

Table 6        IBVS 985 variable stars classification based on Gaia DR2 CAMD

| Name | CAMD classification | Name | CAMD classification |
|---|---|---|---|
| NSV 10577 | Not classified | V3918 Sgr | LPV |
| NSV 10626 | Not classified | V3923 Sgr | LPV |
| NSV 10757 | Not classified | V3924 Sgr | LPV |
| NSV 10837 | Not classified | V3925 Sgr | LPV |
| V3942 Sgr | Not classified | V3926 Sgr | LPV |
| V3934 Sgr | Not LPV | V3927 Sgr | LPV |
| NSV 10741 | SR/IR | V3931 Sgr | LPV |
| NSV 10490 | Probable YSO | V3932 Sgr | LPV |
| NSV 10522 | Probable YSO | V3935 Sgr | LPV |
| V3920 Sgr | Probable YSO | V3936 Sgr | LPV |
| V3921 Sgr | Probable YSO | V3938 Sgr | LPV |
| V3930 Sgr | Probable YSO | V3940 Sgr | LPV |
| V3937 Sgr | Probable YSO | V3943 Sgr | LPV |
| V3939 Sgr | Probable YSO | V3944 Sgr | LPV |
| NSV 10693 | LPV | V3945 Sgr | LPV |
| NSV 10752 | LPV | V3946 Sgr | LPV |
| NSV 10772 | LPV | V3947 Sgr | LPV |
| V3904 Sgr | LPV | V3948 Sgr | LPV |
| V3905 Sgr | LPV | V3949 Sgr | LPV |
| V3908 Sgr | LPV | V3950 Sgr | LPV |

## 2.2. Detailed data analysis

Details of the relevant information available from the bibliographic references, catalogues and databases and the results of our period analysis and assessment are reported in this section for each star.

**NSV 10490**

This star is identified with the infrared counterpart 2MASS J18171849-1734104 (Nesci, 2018) and Gaia DR2/EDR3 source ID 4097207867633543808, that is placed 8 arcsec southwest with respect to the original coordinates, transformed at epoch J2000.0 (see Figure 2).

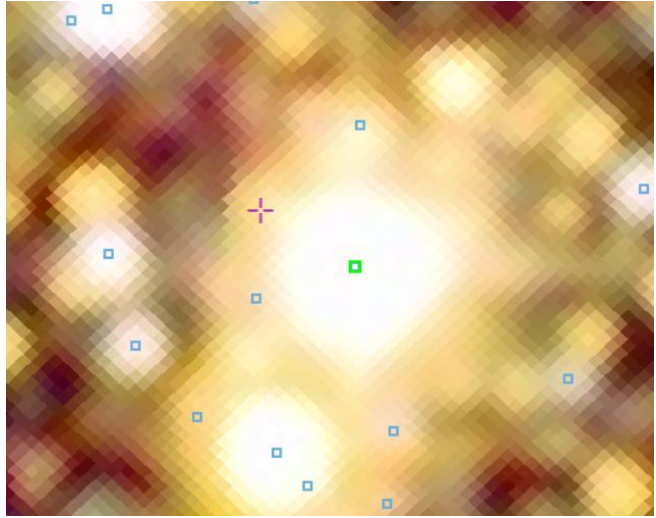

Figure 2    NSV 10490 position (green square) vs. original coordinates (violet cross)

It is classified as an uncertain eclipse binary in the original study. The uncertainty on the type is confirmed by SIMBAD that reports this object as a generic variable star and by AAVSO database which defines it as an uncertain Semiregular based on its color.

The membership of this variable to the Semiregular group is not in accordance with its Gaia DR2 color index Bp-Rp = 6.598 and absolute magnitude $M_G$ = 5.77 (-1.29; +0.80), that place this star in the YSO group of Figure 1.

The Gaia DR2 database (Gaia Collaboration et al. 2018b) does not report any variable star within a radius of 60" from the Gaia EDR3 equatorial coordinates of ID 4097207867633543808.

Also, the ASAS-SN Catalog of Variable Stars II (Jayasinghe et al. 2019) does not report any variable at given Gaia EDR3 coordinates.

We did not perform any period analysis of ASAS-SN photometric data, because no sufficient valid measurements are available.

**NSV 10522**

This star is identified with the infrared counterpart 2MASS J18182855-1725289 (Nesci, 2018). However, based on the original finding chart we identified the variable as the infrared source MSX6C G013.7286-00.8537, that is located 12" southeast from 2MASS J18182855-1725289, and cross referenced also as Gaia DR2/EDR3 4097214842660615808 and 2MASS 18182915-1725379. In turn, MSX6C G013.7286-00.8537 is located 13" west of the original equatorial coordinates, so that it is reported as a different star in SIMBAD (see Figure 3).

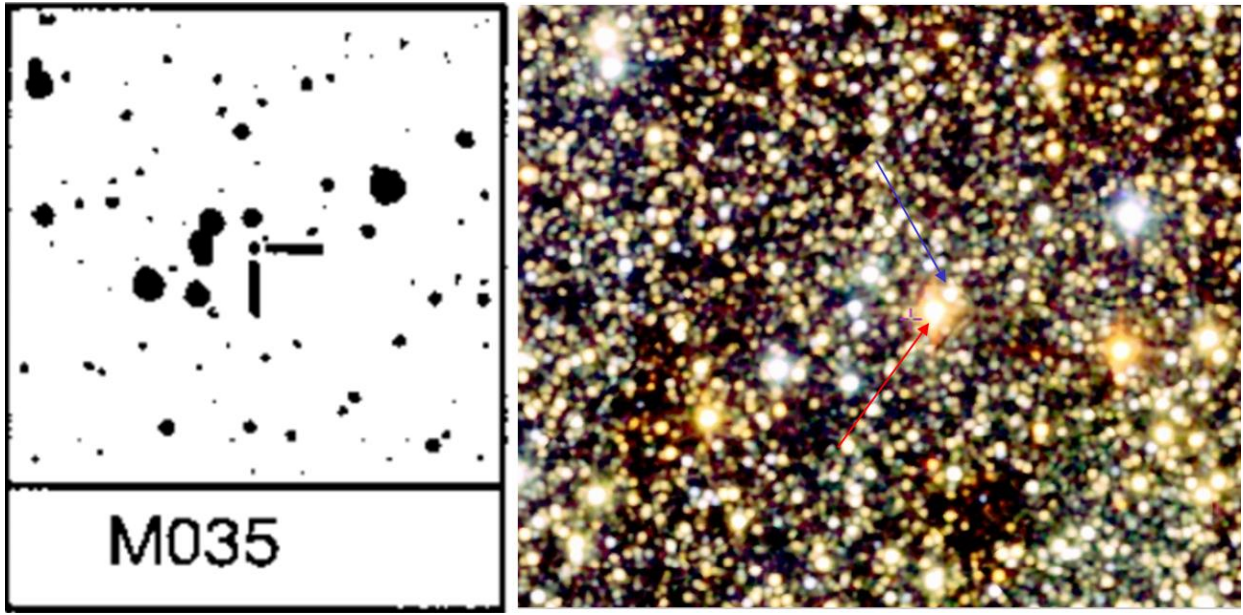

Figure 3   MSX6C G013.7286-00.8537 position (red arrow) vs. 2MASS J18182855-1725289 (blue arrow), original coordinates (violet cross) and original finding chart.

In the AAVSO database the NSV 10522 is classified as an unconfirmed slow irregular variable and cross referenced to the 2MASS J18182915-1725379 and Gaia DR2 4097214842660615808 we identified. No indication on the variable type is provided in the revised characteristics of the variable (Maffei and Tosti 2013), in accordance with SIMBAD, which reports this object as a generic variable star.

The ASAS-SN Catalog of Variable Stars II identifies NSV 10522 with source ASASSN-V J181829.70-172540.0, that is located at a distance of 8" from Gaia DR2/EDR3 coordinates. This source is classified as a Young Stellar Object (YSO) variable, with a mean magnitude V = 15.44 and an amplitude of 0.39 mag, but a different Gaia EDR3 ID 4097214842648260096. The ASAS-SN cross reference to this variable is deemed incorrect because it refers to a star with a mean magnitude G = 19.496, at an angular distance of 10 arcsec from the variable, and JHK magnitudes reported in the database are those of source ID 4097214842660615808. The ASAS-SN classification is consistent with its Gaia DR2 color index Bp-Rp = 5.190 and absolute magnitude $M_G$ = 4.58 (-1.52; +1.31).

It is classified as a large amplitude variable in Gaia DR2 (Mowlavi et al. 2021), with an amplitude $\Delta G$ = 1.11 mag, but it is not reported as candidate YSO in the SVM selection of WISE (Marton et al. 2016).

We performed an analysis of 700 valid ASAS-SN photometric data in the filter V on a time span of 1298 days, applying both Lomb-Scargle and ANOVA methods and despite the YSO type we have identified a potential periodicity. The weighted average of the periods we obtained with the two methods is: 7.475 ± 0.012 days (Figure 4), with a mean average amplitude of 0.10 mag in the V filter. A maximum at epoch 2457842 ± 1 HJD was also identified.

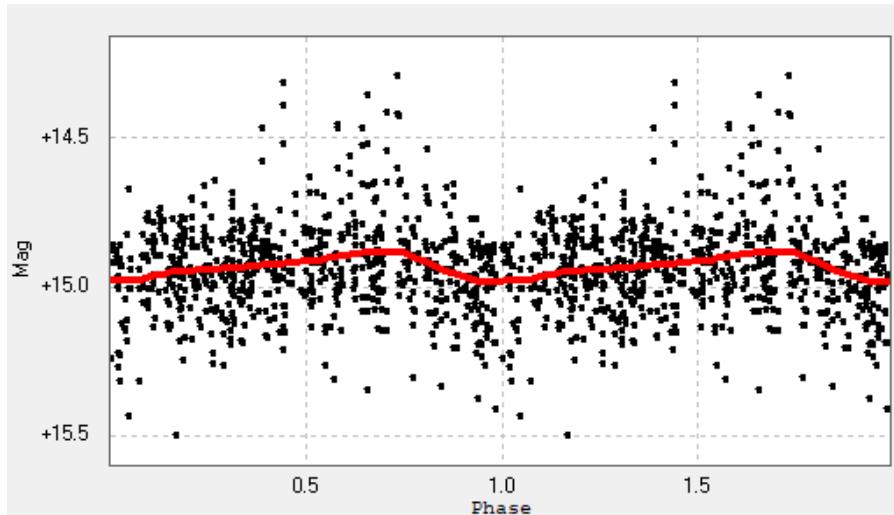

Figure 4   NSV 10522 - V mag vs Phase (7.475 ± 0.012 days period - weighted average)

**NSV 10577 (V5514 Sgr)**

This star is identified with the infrared counterpart 2MASS J18185365-1718281 and Gaia DR2/EDR3 source ID 4097310117901098880, and it is reported as a Mira by SIMBAD and AAVSO, based on the original study which determined a period of 665 days. Even if no distance indication is available from Gaia DR2 and therefore a comparison with the sample of variables of Figure 1 is not possible, the Gaia EDR3 color index Bp-Rp = 6.042 and absolute magnitude $M_G$ = 3.26 (-0.83; +0.88) are consistent with a LPV star. The Gaia EDR3 absolute magnitude has been calculated using the geometric distances available from the Distances to 1.47 billion stars in Gaia EDR3 catalogue (Bailer-Jones et al. 2021).

It is classified as a large amplitude variable in Gaia DR2 with an amplitude $\Delta G$ = 0.94 mag.

It is noted that this variable shows a difference of median magnitude G and color index Bp-Rp between Gaia DR2 and EDR3 databases greater than 1.3, which may indicate a variation in the mean level of luminosity of the star.

The ASAS-SN Catalog of Variable Stars II identifies NSV 10577 with source ASASSN-V J181853.65-171828.1 and classifies it as a Semiregular variable, with a mean magnitude V = 16.98, an amplitude of 0.71 mag and a shorter period of 15 days.

We did not perform any period analysis of ASAS-SN photometric data, because no sufficient valid measurements are available.

**NSV 10626 (V5532 Sgr)**

This star is identified with the infrared counterpart 2MASS J18195847-1731335 and Gaia DR2/EDR3 source ID 4096540567217210368 and is reported as a Mira candidate by SIMBAD and AAVSO, based on the original study which determined an uncertain period of 338 days.

Even if no color index is available from Gaia DR2 and therefore a direct comparison with the sample of variables of Figure 1 is not possible, the Gaia EDR3 color index Bp-Rp = 6.878 and absolute magnitude $M_G$ = 1.41 (-0.93; +1.20) are consistent with a LPV star.

It is classified as a large amplitude variable in Gaia DR2 with an amplitude $\Delta G$ = 0.62 mag.

The Gaia DR2 database does not report any variable star within a radius of 60 arcsec from the Gaia EDR3 coordinates of ID 4096540567217210368.

The ASAS-SN Catalog of Variable Stars II does not report any variable at given Gaia EDR3 coordinates.

We did not perform any period analysis of ASAS-SN photometric data, because no sufficient valid measurements are available.

**NSV 10693**

This star is identified with the infrared counterpart 2MASS J18215979-1634071 and Gaia DR2/EDR3 source ID 4097737660433986816 and is reported as a generic variable star by SIMBAD and AAVSO databases, in accordance with the original study. The Gaia DR2 color index Bp-Rp = 6.529 and absolute magnitude $M_G$ = 3.74 (-1.51; +1.32) are consistent with a LPV (Figure 1). It is classified as a large amplitude variable in Gaia DR2 with an amplitude $\Delta G$ = 0.60 mag. The ASAS-SN Catalog of Variable Stars II identifies NSV 10693 with source ASASSN-V J182159.60-163406 and classifies it as a Semiregular variable, with a mean magnitude V = 13.38, an amplitude of 0.06 mag and a period of 701 days.

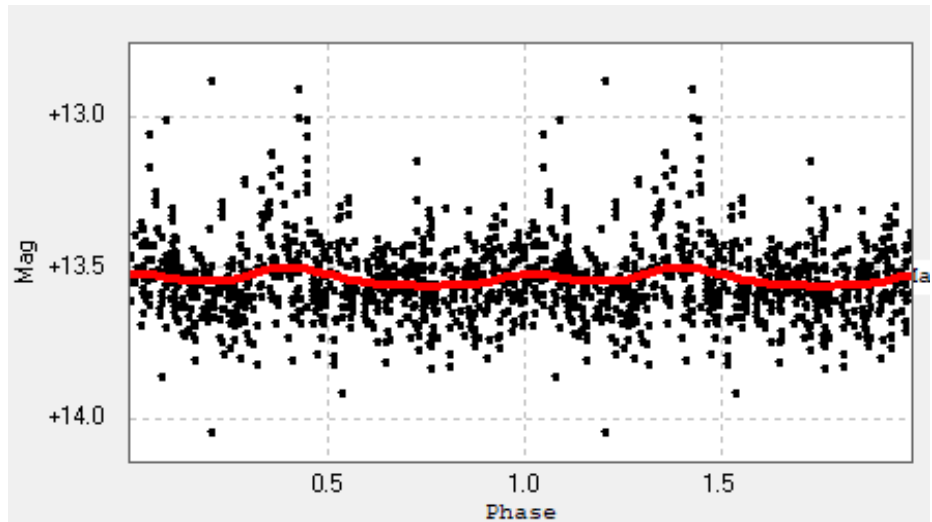

Figure 5     NSV 10693 - V mag vs Phase (17.69 ± 0.06 days period - ANOVA)

Our period analysis, based on 720 valid observations available from ASAS-SN in the V filter and covering a time span of 1298 days was performed applying both Lomb-Scargle and ANOVA methods. We have identified several light curves, with mean amplitude of 0.04÷0.08 mag in filter

V. Potential periods span in a wide range from 18 to 423 days, with shorter periods that show smaller percentage errors. The light curve resulting from the 17.69 days period is shown in Figure 5. We did not find reliable solutions with period around 700 days. We identified a maximum at epoch 2457936 ± 4 HJD.

**NSV 10741**
This star is identified with the infrared counterpart 2MASS J18242539-1703515 (Nesci, 2018), and Gaia DR2/EDR3 source ID 4096948550376111616, that is the northern component of a couple of remarkably close and bright infrared stars. This star was classified as a generic variable by the original study and is reported as an uncertain Mira by AAVSO without any specific reference. The Gaia DR2 color index Bp-Rp = 3.107 and absolute magnitude $M_G$ = 1.35 (-0.77; +0.60) are consistent with a Semiregular or Irregular variable of Figure 1.

The ASAS-SN Catalog of Variable Stars II identifies this variable with source ASASSN-V J182425.20-170356.0, but with a different Gaia EDR3 source ID 4096948550378843264, and classifies it as a generic variable, with a mean magnitude V = 15.10, an amplitude of 0.23 mag and a period of 230 days. The ASAS-SN cross reference is deemed incorrect because refers to the southern component of the pair, 2MASS 18242495-1703556 or Gaia DR2/EDR3 source ID 4096948550378843264. It is noted that this second object is classified as a large amplitude variable in Gaia DR2 with an amplitude ΔG = 0.65 mag. Our period analysis, based on 675 valid observations available from ASAS-SN in the V filter and covering a time span of 1298 days, was performed applying both Lomb-Scargle and ANOVA methods. The weighted average of the periods we obtained with the two methods is 227 ± 7 days (Figure 6), that confirms the ASAS-SN result and is consistent with a Semiregular variable type. We identified a maximum at epoch 2457881 ± 7 HJD.

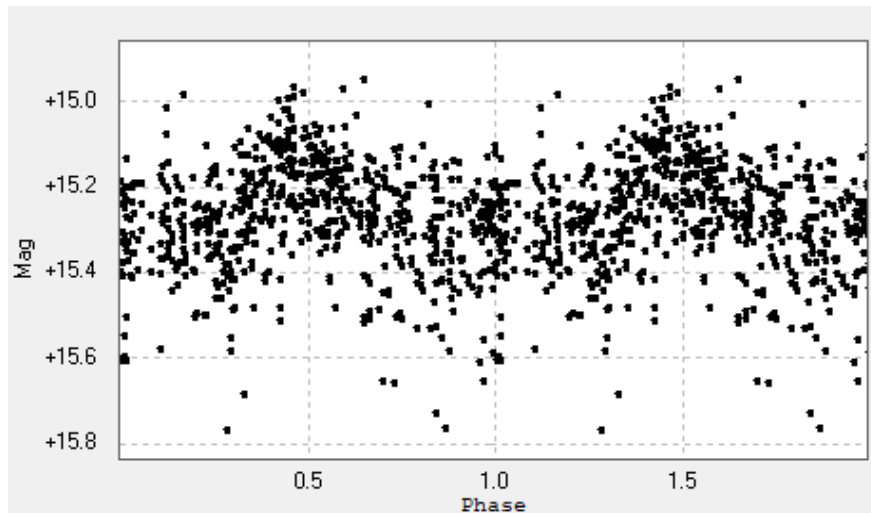

Figure 6    NSV 10741 - V mag vs Phase (227 ± 7 days period - weighted average)

**NSV 10752 (V5536 Sgr)**
This star is identified with the infrared counterpart 2MASS J18250499-1703578 and Gaia DR2/EDR3 source ID 4096900966474840320 and it is reported as a Mira by both SIMBAD and AAVSO database in accordance with the original study, which found a period of 206 days. This classification is consistent with its Gaia DR2 color index Bp-Rp = 4.672 and absolute magnitude $M_G = 1.80 \pm 0.89$, which place this variable in the LPV group of Figure 1.

It is classified as a large amplitude variable in Gaia DR2, with an amplitude $\Delta G = 0.56$ mag.

The Bochum Galactic Disk Survey II refers to this variable as source ID GDS J1825050-170358, with a median light curve magnitude i = 15.90 and a maximum amplitude of 3.13 mag.

It is also reported as a WISE J182505.00-170357.8 candidate YSO.

The ASAS-SN Catalog of Variable Stars II does not report any variable at given Gaia EDR3 coordinates.

We did not perform any period analysis of ASAS-SN photometric data, because no sufficient valid measurements are available.

**NSV 10757**

This is a faint star reported in the original study as a generic variable, with no period defined, which achieved in the infrared photographic observations a maximum of luminosity of 16.2 mag and a minimum fainter than 17.2 mag and is identified with the infrared counterpart 2MASS J18250968-1610350 (Nesci, 2018).

We could not identify this object, that is located about 5" west from the original coordinates, with a common Gaia DR2/EDR3 source ID. We cross matched it with Gaia DR2 4097858602406503552 (G = 15.755) and Gaia EDR3 4097858606748077696 (G = 17.134). Each ID is present only in its specific release and cannot be resolved in the other one.

Because the angular separation between the two sources is < 0.2" and the difference of their color index Bp-Rp is < 0.1, we believe that is highly likely that the two different IDs correspond to the same object, despite of the significant difference in the mean magnitude.

Both IDs have photometric characteristics consistent with a LPV star and based on its color and amplitude the AAVSO database classifies this an uncertain Semiregular.

It is also reported as a WISE J182509.68-161035.1 candidate YSO.

The source 4097858602406503552 is classified as a large amplitude variable in Gaia DR2, with an amplitude $\Delta G = 0.19$ mag.

The ASAS-SN Catalog of Variable Stars II does not report any variable in a radius of 10" cantered at the coordinates of star 2MASS J18250968-1610350. Moreover, using 2MASS or Gaia DR2 ID or Gaia EDR3 ID coordinates, the ASAS-SN returns the light curve of the same object. We did not perform any period analysis of the ASAS-SN photometric data for this object because no sufficient valid measurements are available.

**NSV 10772**

This star is identified with the infrared counterpart 2MASS J18254743-1611475 (Nesci, 2018) and Gaia DR2/EDR3 source ID 4097856270277660032. It is a faint object, with a maximum of 16.2 mag and an amplitude of 0.8 mag, with no type or period defined in the original study. As well, the INTEGRAL OMC catalogue reports this object as a generic variable (Alfonso-Garzón et al. 2012). The AAVSO database reports this variable as an uncertain Semiregular based on its color and amplitude. This classification is consistent with the Gaia DR2 color index Bp-Rp =

6.354 and absolute magnitude $M_G$ = 1.30 (-1.01; +1.03), which place this object in the LPV group of Figure 1. It is also reported as a WISE J182547.43-161147.6 candidate YSO.

The ASAS-SN Catalog of Variable Stars II reports this star as ASASSN-V J182547.10-161144.0 and classifies this object as a slow irregular variable, with a mean magnitude V = 12.09 and an amplitude of 0.75 mag.

We performed a period analysis, using 724 valid observations available from ASAS-SN in the V filter and covering a time span of 1298 days applying both Lomb-Scargle and ANOVA methods. We have identified several potential periods which span in a wide range from 15 to 700 days with mean amplitude of the light curve not exceeding 0.11 mag; Figure 7 shows the light curve with lower error in percentage. We identified a maximum at epoch 2459073 ± 3 HJD.

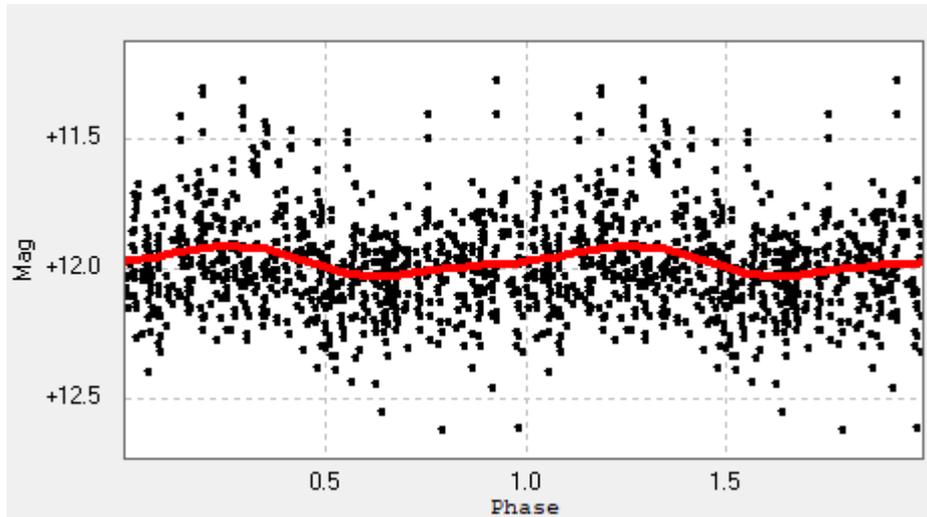

Figure 7   NSV 10772 - V mag vs Phase (36.7 ± 0.1 days period - Lomb-Scargle)

**NSV 10837 (V5538 Sgr)**

This variable achieved a maximum of 16.1 and was fainter than 17.4 mag in the original infrared photographic system. The AAVSO database reports it as a Semiregular (SRA), in accordance with the original study, which determined a period of 416 days. SIMBAD refers to this star as a Mira and cross matches it as 2MASS J18275407-1621276 and Gaia DR2 source ID 4097091010218371840.

We could not identify this variable with a unique Gaia ID because the Gaia database reports as more likely counterpart the Gaia DR2 4097091010218371840 (G = 14.799) or Gaia EDR3 4097091005854942848 (G = 16.106). Each ID is present only in its specific release and cannot be resolved in the other one.

Because the angular separation between the two sources is 4 mas and the difference of their color index Bp-Rp is < 0.3, we believe that is highly likely that the two different IDs correspond to the same object, despite of the significant difference in the mean magnitude.

The source 4097091010218371840 is classified as a large amplitude variable in Gaia DR2, with an amplitude ΔG = 0.14 mag.

The Bochum Galactic Disk Survey II refers to this variable as source ID GDS J1827548-162128, with a median light curve magnitude i = 15.36 and a maximum amplitude of 1.72 mag.

It is also reported as a WISE J182754.06-162127.6 candidate YSO.

Using both the coordinates of Gaia DR2 4097091010218371840 and Gaia EDR3 4097091005854942848, the ASAS-SN Catalog of Variable Stars II identifies this star always with the same object, ASASSN-V J182754.07-162127.6, and classifies it a s a generic variable, with an undefined period, a mean magnitude V = 15.75 and an amplitude of 0.29 mag.

We performed a period analysis, using 404 valid observations available from ASAS-SN in the V and g filters, and covering a time span of 1298 days, applying both Lomb-Scargle and ANOVA methods but not reliable period was found for this variable.

**V3904 Sgr**

This star is identified with the infrared counterpart 2MASS J18110608-1613039 and Gaia DR2/EDR3 4145671007298104960. It is reported by SIMBAD as a Mira candidate and by AAVSO as a Mira, in accordance with the original study, which determined a period of 360 days. This classification is consistent with the Gaia DR2 color index Bp-Rp = 7.060 and absolute magnitude $M_G$ = 3.06 (-1.63; +1.19) which place this object in the LPV group of Figure 1. The source 4145671007298104960 is classified as a large amplitude variable in Gaia DR2, with an amplitude ΔG = 0.47 mag.

The ASAS-SN Catalog of Variable Stars II identifies this star as ASASSN-V J181106.09-161304.0 and classifies it as a Semiregular, with a mean magnitude V = 15.43, an amplitude of 0.16 mag and a period of 57 days.

We performed a period analysis, using 632 valid observations in the filter V and 404 in the filter g filter, covering a time span of 1298 days, available from ASAS-SN. Applying both Lomb-Scargle and ANOVA methods we found weak evidence of periodicity in the light curve of the variable in the filter V. A potential period of 29.6 ± 0.4 days was identified with the ANOVA method in the filter g (Figure 8). No reliable solution was found around 60 days.

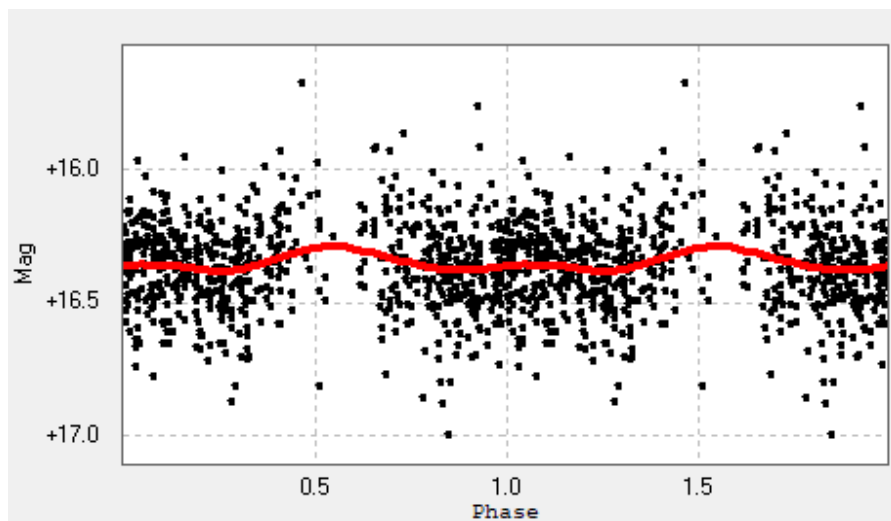

Figure 8    V3904 Sgr - g mag vs Phase (29.6 ± 0.4 days period - ANOVA)

**V3905 Sgr**

This star is identified with the infrared counterpart 2MASS J18111338-1603290 and Gaia DR2/EDR3 source ID 4145677569997394688, and classified as a large amplitude variable in Gaia DR2, with an amplitude ΔG = 0.51 mag. The classification reported by AAVSO reflects the Semiregular (SRa) type defined in the original study, which determined a period of 328 days. The Gaia DR2 color index Bp-Rp = 5.288 and absolute magnitude $M_G$ = 3.74 (-1.29; +1.31) are compatible with the LPV group of Figure 1.

This variable shows also short-time sudden increase of magnitude, which occurs at phase = 0.92, with an amplitude of 0.7 mag (I-N hypersensitized+RG5) and a duration of 3.0 days (Maffei and Tosti, 1995). It is also reported as a WISE J181113.38-160329.1 candidate YSO. At Gaia EDR3 equatorial coordinates, the ASAS-SN Catalog of Variable Stars II does not identify any variable star. We did not perform any period analysis of ASAS-SN photometric data, because no sufficient valid measurements are available.

**V3908 Sgr**

This a M8 spectral type star (Stephenson, 1992), identified with the infrared counterpart 2MASS J18133473-1613495 and Gaia DR2/EDR3 source ID 4145615030968253952. This classification is consistent with the Gaia DR2 color index Bp-Rp = 5.922 and absolute magnitude $M_G$ = 0.75 (-0.58; +0.46), which place this object in the LPV group of Figure 1.

The Gaia DR2 database classifies this star as LPV Candidate with a period of 269 ± 17 days (Mowlavi et al. 2018) and an amplitude ΔG = 0.44 mag.

The ASAS-SN Catalog of Variable Stars II refers to this star as ASASSN-V J181334.82-161349.7 and classifies it as a Mira, with a mean magnitude V = 14.39, an amplitude of 4.44 mag and a period of 281 days.

We performed an analysis of 584 ASAS-SN photometric data in the filter V on a time span of 1298 days, applying both Lomb-Scargle and ANOVA methods. The weighted average of the periods we obtained with the two methods is 281 ± 1 days (Figure 9), that confirms the previous results. A maximum at epoch 2457863 ± 1 HJD was also identified.

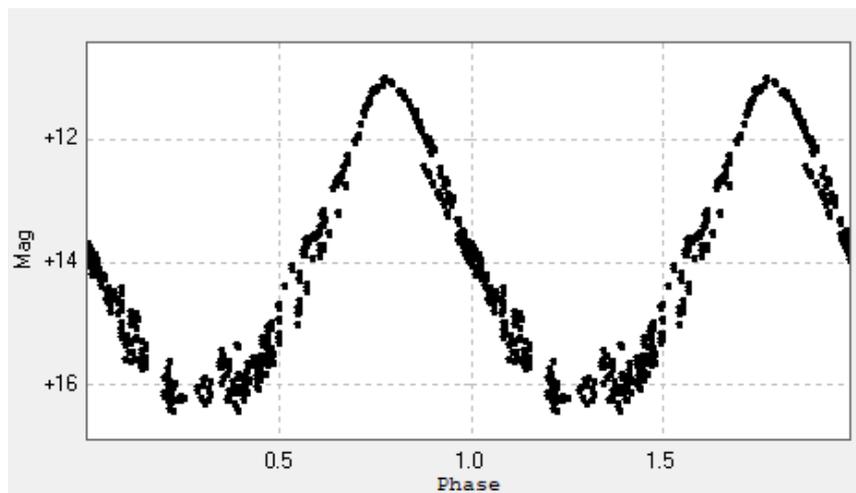

Figure 9    V3908 Sgr - V mag vs Phase (281 ± 1 day period - weighted average)

**V3918 Sgr**

Due to a misprint in the initial IBVS 985, the equatorial coordinates currently reported in the SIMBAD database for this variable are not correct. This star is identified with the infrared counterpart 2MASS J18290441-1353350 (Nesci, 2018) and Gaia DR2/EDR3 source ID 4104382730793346688. The R.A. = 18h 29m 04.41552s and Dec. = -13° 53' 35.0268" (J2000) place this object in the Scutum constellation, despite of its designation V3918 Sgr.

SIMBAD and AAVSO database report this variable as a Mira candidate and Mira respectively, in accordance with the original classification, that determined a period of 384 days.

We identified as Gaia DR2/EDR3 counterpart of this variable the source ID 4104382730793346688. This source is classified as a large amplitude variable in Gaia DR2 catalogue, with an amplitude $\Delta G = 0.42$ mag.

The Bochum Galactic Disk Survey II refers to this variable as source ID GDS J1829044-135334, with a median light curve magnitude i = 14.50 and a maximum amplitude of 2.47 mag.

The membership of this object to the Mira group is consistent with its color index Bp-Rp = 6.637 and absolute magnitude $M_G = 1.60$ (-1.42; +1.32).

The ASAS-SN Catalog of Variable Stars II does not report any variable in a radius of 10" from the Gaia EDR3 equatorial coordinates of source ID 4104382730793346688.

We did not perform any period analysis of ASAS-SN photometric data, because no sufficient valid measurements are available.

**V3920 Sgr**

This star is identified with the infrared counterpart 2MASS J18173387-1732054 and Gaia DR2/EDR3 source ID 4097209551260714880. SIMBAD and AAVSO database report this variable as an LPV candidate or uncertain Semiregular with unknown period, in accordance with the original study.

The Gaia color index Bp-Rp = 6.335 and absolute magnitude $M_G = 4.75$ (-0.76; +0.57) place this variable in the YSO group of Figure 1. This star is classified as a large amplitude variable in Gaia DR2 catalogue, with an amplitude $\Delta G = 0.77$ mag.

The ASAS-SN Catalog of Variable Stars II incorrectly cross matches V3920 Sgr to source ASASSN-V J181732.51-173158.8, 2MASS J18173251-1731587 and different Gaia EDR3 4097209551260717312. This source is classified as a generic variable, with a mean magnitude V = 16.54 and amplitude of 0.57 mag and is placed 21" from Gaia DR2/EDR3 source ID 4097209551260714880 (see Figure 10).

We did not perform any period analysis of ASAS-SN photometric data of source that matches Gaia DR2/EDR3 4097209551260714880 coordinates because no sufficient valid measurements are available.

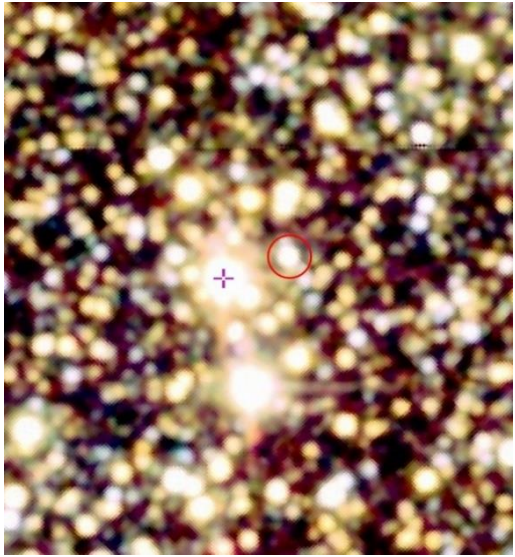

Figure 10    V3920 Sgr correct identification (cross) vs. ASAS-SN source (red circle)

**V3921 Sgr**

The variable star V3921 Sgr is identified with the infrared counterpart 2MASS J18173623-1741404 and Gaia DR2/EDR3 source ID 4097202748032445312, and classified as a Mira candidate, based on the original study, which determined a period of 531 days.

This classification is not consistent with the Gaia DR2 color index Bp-Rp = 6.516 and absolute magnitude $M_G$ = 5.67 (-1.29; +0.81) which place this star in the YSO group of Figure 1.

This source is classified as a large amplitude variable in Gaia DR2 catalogue, with an amplitude $\Delta G$ = 0.97 mag.

The ASAS-SN Catalog of Variable Stars II does not report any variable in a radius of 10" from the Gaia EDR3 equatorial coordinates of source ID 4097202748032445312.

We did not perform any period analysis of ASAS-SN photometric data, because no sufficient valid measurements are available.

**V3923 Sgr**

This star is identified with the infrared counterpart J2MASS 18184943-1743528 and Gaia DR2/EDR3 source ID 4097188106476834816. The original study classified this variable as an uncertain Mira, with an undefined period. This classification is consistent with the colour index Bp-Rp = 5.284 and absolute magnitude $M_G$ = 3.00 (-1.15; +1.13), which collocate this variable in the LPV group of Figure 1.

This source is classified as a large amplitude variable in the Gaia DR2 catalogue, with an amplitude $\Delta G$ = 0.91 mag.

The ASAS-SN Catalog of Variable Stars II identifies this star as ASASSN-V J181849.44-174352.8and classifies it as Semiregular, with a mean magnitude V = 16.95, an amplitude of 0.61 mag and a period of 8 days.

We did not perform any analysis of the ASAS-SN photometric data due to insufficient valid data available.

**V3924 Sgr**

This star is identified with the infrared counterpart 2MASS J18192768-1745477 and Gaia DR2/EDR3 4096442298393113856 and reported by SIMBAD as a Mira candidate and by AAVSO as a Mira, in accordance with the original study, which determined a period of 372 days. This classification is consistent with the Gaia DR2 color index Bp-Rp = 5.919 and absolute magnitude $M_G$ = 2.74 (-1.05; +1.01), which place this object in the LPV group of Figure 1.

It is classified as a large amplitude variable in the Gaia DR2 catalogue, with an amplitude $\Delta G$ = 0.87 mag and is reported as a WISE J181927.69-174547.4 candidate YSO.

The ASAS-SN Catalog of Variable Stars II identifies this star as ASASSN-V J181927.69-174547.7 and classifies it as a Semiregular, with a mean magnitude V = 14.97, an amplitude of 0.12 mag and a period of 6 days.

Our period analysis, based on 865 valid observations available from ASAS-SN in the V filter and covering a time span of 1298 days, applying the Lomb-Scargle and ANOVA methods, did not highlight any reliable period.

**V3925 Sgr**

This star is identified with the infrared counterpart 2MASS J18193309-1746442 and Gaia DR2/EDR3 source ID 4096439274736278272. The Gaia DR2 database classifies this star as LPV Candidate with a period of 148 ± 8 days, in accordance with the original classification and period of 151.5 days.

The Gaia DR2 classification is consistent with the colour index Bp-Rp = 5.995 and absolute magnitude $M_G$ = 2.82 (-1.25; +0.90), which collocate this variable in the LPV group of Figure 1.

It is also classified as a large amplitude variable in Gaia DR2 catalogue, with an amplitude $\Delta G$ = 0.73 and reported as a WISE J181933.08-174644.2 candidate YSO.

The ASAS-SN Catalog of Variable Stars II does not report any variable in a radius of 10" from the Gaia EDR3 equatorial coordinates of source ID 4096439274736278272.

We did not perform any analysis of the ASAS-SN photometric data due to insufficient valid data available.

**V3926 Sgr**

This star is identified with the infrared counterpart 2MASS J18193414-1748395 and Gaia DR2/EDR3 source ID 4096438617497438720. SIMBAD and AAVSO database report this variable as a Mira, in accordance with the original study, which determined a period of 290 days. This classification is consistent with the color index Bp-Rp = 5.981 and absolute magnitude $M_G$ = 3.49 (-1.20; +0.80), which collocate this variable in the LPV group of Figure 1.

This source is classified as a large amplitude variable in Gaia DR2 catalogue, with an amplitude $\Delta G$ = 0.54 mag.

The ASAS-SN Catalog of Variable Stars II identifies this star, ASASSN-V J181934.15-174839.6, as a Semiregular, a mean magnitude V = 15.9, an amplitude of 0.31 mag and a period of 274 days.

Our period analysis, based on a limited number of valid observations available from ASAS-SN in the V and g filters, applying the Lomb-Scargle and ANOVA methods, did not highlight any reliable period.

**V3927 Sgr**

This star is identified with the infrared counterpart 2MASS J18195744-1743099 and Gaia DR2/EDR3 source ID 4096441198881091328. The AAVSO database reports this variable as a Semiregular with a period of 218 days, in accordance with the original study, whilst SIMBAD refers to it as a Mira candidate. Both classifications are consistent with the Gia DR2 color index Bp-Rp = 5.719 and absolute magnitude $M_G$ = 0.84 (-1.11; +1.00), which collocate this variable in the LPV group of Figure 1.

This source is classified as a large amplitude variable in the Gaia DR2 catalogue, with an amplitude $\Delta G$ = 0.50 mag.

The ASAS-SN Catalog of Variable Stars II identifies this star as ASASSN-V J181957.44-174310.0 and classifies it as a Semiregular, a mean magnitude V = 14.08, an amplitude of 0.11 mag and a period of 257 days.

Our period analysis was performed using 1103 (filter V) valid measurements available from ASAS-SN, covering a time span of 1298 days and applying both Lomb-Scargle and ANOVA methods. We have identified several potential periods which span in a wide range from 7 to 219 days, with mean amplitude of the light curve not exceeding 0.07 mag, with shorter periods that show smaller percentage errors. No reliable solution was found for period around 257 days.

**V3930 Sgr**

This star is identified with the infrared counterpart 2MASS J18210828-1706231 and Gaia DR2/EDR3 source ID 4097329329300561280. The SIMBAD and AAVSO databases report this variable as a Mira, in accordance with the original study, which determined a period of 448 days. This classification is not consistent with the Gaia DR2 color index Bp-Rp = 5.642 and absolute magnitude $M_G$ = 4.36 (-1.49; +1.58), which collocate this variable in the YSO group of Figure 1. This source is classified as a large amplitude variable in the Gaia DR2 catalogue, with an amplitude $\Delta G$ = 0.24 mag and reported as a WISE J182108.29-170623.0 candidate YSO.

The ASAS-SN Catalog of Variable Stars II identifies this star as ASASSN-V J182108.29-170623.1 and classifies it as a Semiregular, a mean magnitude V = 17.16, an amplitude of 0.38 mag and a period of 7 days.

The ASAS-SN cross-reference ID is deemed incorrect because we could not find this source within a circle of 60" centered on the equatorial coordinates of V3930 Sgr, and photometric data reported in the database are those of source ID 4097329329300561280.

We did not perform any analysis of the ASAS-SN photometric data due to insufficient valid data available.

**V3931 Sgr**

This star is identified with the infrared counterpart 2MASS J18212174-1642586 and Gaia DR2/EDR3 source ID 4097359462791432704. The SIMBAD and AAVSO database report this star as Mira, in accordance with the original study, which found a period of 441 days. This classification is consistent with the Gaia DR2 color index Bp-Rp = 7.212 and absolute magnitude $M_G$ = 0.98 (-1.17; +1.16), which collocate this star in the LPV group shown in Figure 1.

This source is classified as a large amplitude variable in Gaia DR2 catalogue, with an amplitude ΔG = 0.46 mag and the Bochum Galactic Disk Survey II refers to this variable as source ID GDS J1821217-164259, with a median light curve magnitude i = 15.12 and a maximum amplitude of 3.35 mag. This variable shows also short-time, sudden increase of magnitude, which occurs at phase = 0.87, with an amplitude of 1.0 mag, in the infrared photographic filter I-N hypersensitized+ RG5 and a duration of 2.0 days (Maffei and Tosti, 1995).

The ASAS-SN Catalog of Variable Stars II does not report any variable star in a radius of 10" of Gaia EDR3 source ID 4097359462791432704 coordinates. We did not perform any analysis of the ASAS-SN photometric data due to insufficient valid data available.

**V3932 Sgr**

This star is identified with the infrared counterpart 2MASS J18214251-1732069 and Gaia DR2/EDR3 source ID 4096532213423534336. SIMBAD and AAVSO database report this star as Mira, in accordance with the original study, which found a period of 205 days.

The membership of this star to the Mira group is consistent with the colour index Bp-Rp = 5.854 and the absolute magnitude $M_G$ = 0.15 (-0.86; +0.87), which collocate this variable in the LPV group of Figure 1. This source is classified as a large amplitude variable in Gaia DR2 catalogue, with an amplitude ΔG = 0.40 mag.

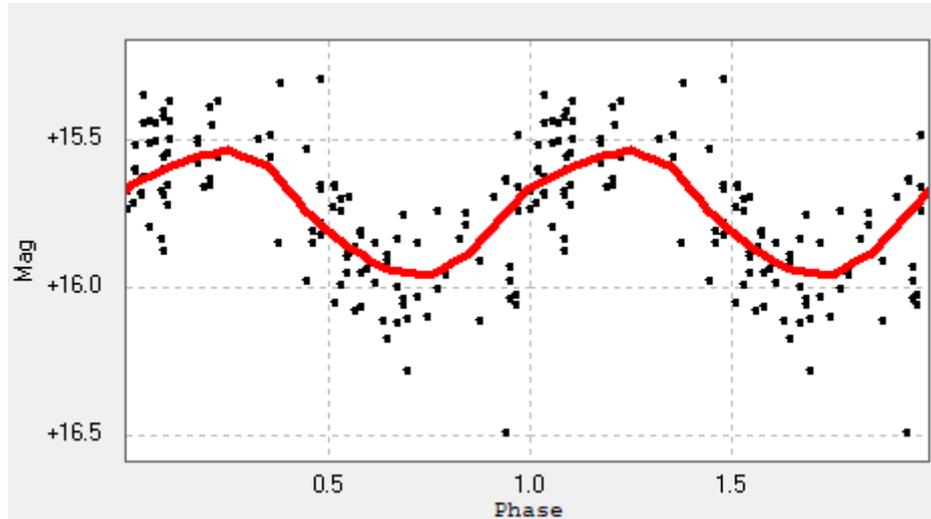

Figure 11    V3932 Sgr - V mag vs Phase (58.4 ± 0.3 days period - Lomb-Scargle)

The ASAS-SN Catalog of Variable Stars II identifies this star as ASASSN-V J182142.53-173207.5 and classifies this variable as Semiregular, with a mean magnitude V = 16.16, an amplitude of 1.5 mag and a period of 201 days.

Our period analysis, based on 117 valid observations available from ASAS-SN in the V filter and covering a time span of 1046 days, applying the Lomb-Scargle methods, highlighted a period of 58.4 ± 0.3 days, with a mean magnitude amplitude of 0.42 mag (Figure 11). We did not find any reliable solution with period around 200 days.

**V3934 Sgr**

This star is identified with the infrared counterpart J18230629-1721099 and Gaia DR2/EDR3 source ID 4096561664098938368. The AAVSO database reports this variable as a Semiregular (SRA), in accordance with the original study, which determined a period of 366 days, whilst the SIMBAD database refers to this variable as a LPV candidate.

The Gaia DR2 color index Bp-Rp = 2.565 and absolute magnitude $M_G$ = 2.37 (-0.37; +0.32) place this variable out of the LPV group of Figure 1.

The Bochum Galactic Disk Survey II refers to this variable as source ID GDS J1823063-172112, with a median light curve magnitude r = 14.80, i = 13.73 and a maximum amplitude of 0.47 mag. The ASAS-SN Catalog of Variable Stars II identifies this star, ASASSN-V J182306.30-172109.9, with a different Gaia EDR3 source ID 4096561659747632000, and classifies it as a generic variable, with a mean magnitude V = 14.78, an amplitude of 0.15 mag and a period of 181 days. The ASAS-SN cross-reference ID is deemed incorrect because source ID 4096561659747632000 is located 2.3" from the equatorial coordinates of V3934 Sgr.

Our period analysis was performed using 699 valid measurements in the filter V available from ASAS-SN, covering a time span of 1298 days and applying both Lomb-Scargle and ANOVA methods. We have identified several potential periods which span in a wide range from 29 to 188 days, with mean amplitude of the light curve not exceeding 0.07 mag.

**V3935 Sgr**

This star is identified with the infrared counterpart 2MASS J18240935-1703399 and Gaia DR2/EDR3 source ID 4096942885361078912. The SIMBAD and AAVSO databases report this star as LPV candidate and Mira respectively, in accordance with the original study, which found a period of 395 days.

The membership of this star to the Mira group is consistent with the colour index Bp-Rp = 5.560 and the absolute magnitude $M_G$ = 2.56 ± 1.07, which collocate this variable in the LPV group of Figure 1.

It is classified as a large amplitude variable in the Gaia DR2 catalogue, with an amplitude ΔG = 0.49 mag, and is reported as a WISE J182409.35-170339.8 candidate YSO.

The ASAS-SN Catalog of Variable Stars II identifies this star as ASASSN-V J182409.36-170339.9, and classifies this variable as Semiregular, with a mean magnitude V = 15.06, an amplitude of 0.16 mag and a period of 15 days.

Our period analysis, based on 690 valid observations available from ASAS-SN in the V filter and covering a time span of 1298 days, applying the Lomb-Scargle algorithm did not find any reliable solution for the period.

**V3936 Sgr**

This star is identified with the infrared counterpart 2MASS J18241836-1707212 and Gaia DR2/EDR3 source ID 4096941751489519616. SIMBAD and AAVSO database report this variable as a Mira in accordance with the original study, which determined a period of 369 days.

The membership of this star to the Mira group is consistent with the colour index Bp-Rp = 6.276 and the absolute magnitude $M_G$ = 2.44 (-1.33; +1.23), which collocate this variable in the LPV group of Figure 1.

It is classified as a large amplitude variable in Gaia DR2 catalogue, with an amplitude ΔG = 0.29 mag and the Bochum Galactic Disk Survey II refers to this variable as source ID GDS J1824183-170721, with a median light curve magnitude i = 15.36 and a maximum amplitude of 1.19 mag. The ASAS-SN Catalog of Variable Stars II identifies this star as ASASSN-V J182418.37-170721.3 and classifies it as Semiregular, with a mean magnitude V = 15.74, an amplitude of 0.44 mag and a period of 44 days.

Our period analysis, based on 129 valid observations available from ASAS-SN in the V filter and covering a time span of 1201 days, applying the ANOVA method, highlighted a period of 319 ± 47 days, consistent with the original value (Figure 12).

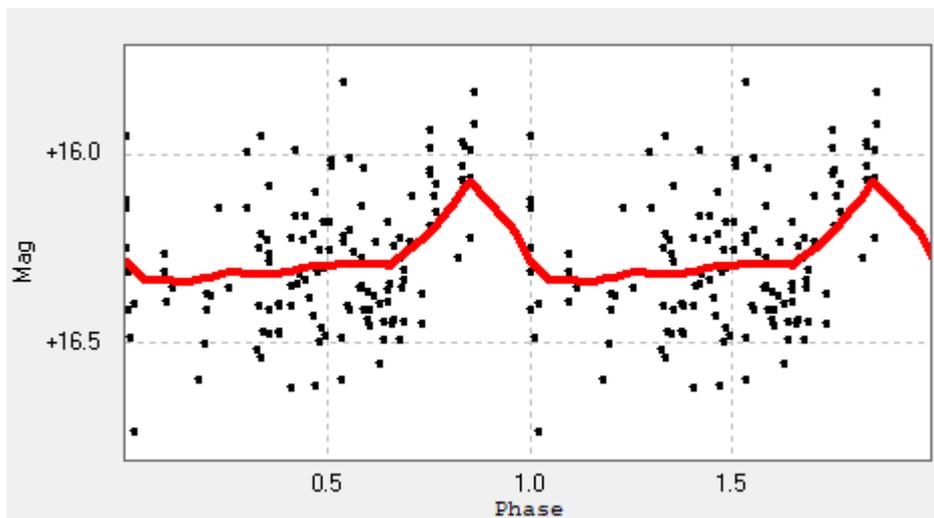

Figure 12    V3936 Sgr - V mag vs Phase (319 ± 47 days period - ANOVA)

**V3937 Sgr**

This a M5 spectral type star (Stephenson, 1992), identified with the infrared counterpart 2MASS J18242712-1610054 and Gaia DR2/EDR3 source ID 4097777444701491584. SIMBAD and AAVSO database report this variable as a Mira, in accordance with the original study that determined a period of 606 days.

This classification is not consistent with the color index Bp-Rp = 7.027 and absolute magnitude $M_G$ = 5.49 (-2.07; +1.11), which collocate this variable in the YSO group of Figure 1.

It is classified as a large amplitude variable in the Gaia DR2 catalogue, with an amplitude ΔG = 0.73.

The ASAS-SN Catalog of Variable Stars II identifies this star as ASASSN-V J182427.13-161005.5, and classifies it as Semiregular, with a mean magnitude V = 16.61, an amplitude of 0.64 mag and a period of 13 days.

We did not perform any analysis of the ASAS-SN photometric data due to insufficient valid data available.

**V3938 Sgr**

This star is identified with the infrared counterpart 2MASS J18243941-1705018 and Gaia DR2/EDR3 source ID 4096947592644959232. The SIMBAD and AAVSO database report this variable as a Mira, based on the original study, which found a period of 297 days. This classification is consistent with the Gaia DR2 color index Bp-Rp = 6.814 and absolute magnitude $M_G$ = 1.70 (-1.21; +1.11), which collocate this variable in the LPV group of Figure 1.

This object is classified as a large amplitude variable in Gaia DR2 catalogue, with an amplitude of 0.67 mag and reported as a WISE J182439.42-170501.9 candidate YSO.

The Bochum Galactic Disk Survey II refers to this variable as source ID GDS J1824394-170502, with a median light curve magnitude r = 16.69, i = 13.67 and a maximum amplitude of 2.69 mag. The ASAS-SN Catalog of Variable Stars II does not report any light curve for this variable. We did not perform any analysis of the ASAS-SN photometric data due to insufficient valid data available.

**V3939 Sgr**

This star is identified with the infrared counterpart 2MASS J18244234-1609067 and Gaia DR2/EDR3 source ID 4097871182403713152. SIMBAD and AAVSO database report this variable as a Mira candidate, in accordance with the original study, which determined a period of 292 days. This classification is not consistent with the Gaia DR2 color index Bp-Rp = 6.350 and absolute magnitude $M_G$ = 5.94 (-0.69; +0.53), which collocate this variable in the YSO group of Figure 1. It is classified as a large amplitude variable in the Gaia DR2 catalogue, with an amplitude $\Delta G$ = 0.70 mag. The ASAS-SN Catalog of Variable Stars II does not report any light curve for this variable. We did not perform any analysis of the ASAS-SN photometric data due to insufficient valid data available.

**V3940 Sgr**

This star is identified with the infrared counterpart 2MASS J18250931-1636201 and Gaia DR2/EDR3 source ID 4097000571017265408. The SIMBAD and AAVSO database report this variable as a Mira, based on the original study, which found a period of 449 days. This classification is consistent with the Gaia DR2 color index Bp-Rp = 7.041 and absolute magnitude $M_G$ = 2.79 (-2.26; +1.27), which collocate this variable in the LPV group of Figure 1.

This star is classified as a large amplitude variable with an amplitude $\Delta G$ = 0.34 mag and the Bochum Galactic Disk Survey II refers to this variable as source ID GDS J1825093-163620, with a median light curve magnitude r = 16.67, i = 12.32 and a maximum amplitude of 2.84 mag.

The ASAS-SN Catalog of Variable Stars II refers to this star as ASASSN-V J182509.31-163620.1 and classifies it as a Semiregular, with a mean magnitude V = 15.78, an amplitude of 0.53 mag and a period of 449 days.

We performed an analysis of 617 ASAS-SN photometric data in both filter V and g, on a time span of 1073 days, applying both Lomb-Scargle and ANOVA methods. The weighted average of the four periods we obtained with the two methods applied two both filters is 444 ± 24 days (Figure 13), in accordance with original and the ASAS results, and with a mean average amplitude of 0.25 mag in the V filter and 0.23 mag in the g filter.

A maximum at epoch 2458961 ± 6 HJD was also identified.

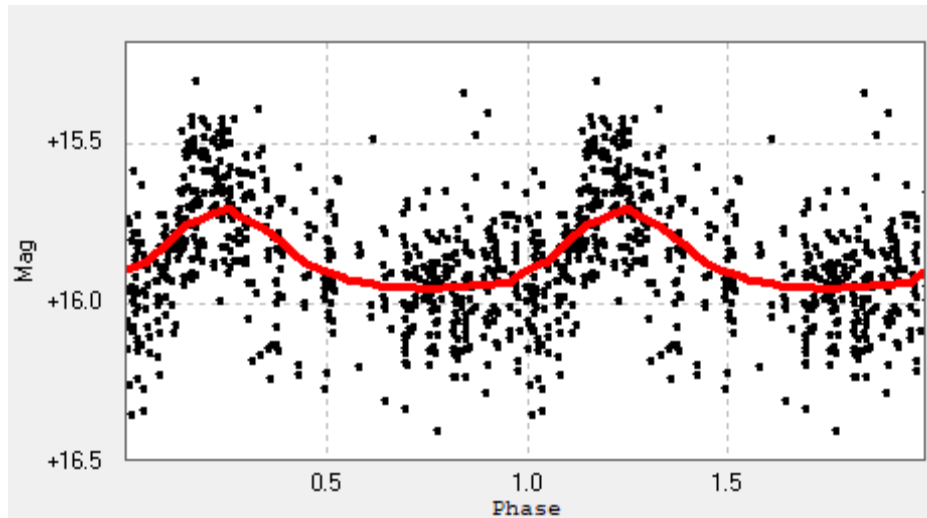

Figure 13    V3940 Sgr - V mag vs Phase (444 ± 24 days period – weighted average)

**V3942 Sgr**

This star is identified with Gaia DR2 source ID 4097002430790019712 (G=12.261), the Gaia EDR3 source ID 4097002430790019840 (G=14.172) and the infrared counterpart 2MASS J18253922-1633460. Because the angular distance between the two Gaia sources is 3 mas it is highly likely that the two IDs correspond to a unique source, despite the significant difference between the mean magnitudes. Each ID is present only in its specific release and cannot be resolved in the other one.

SIMBAD and AAVSO database report this variable as a Mira, in accordance with the original study, which determined a period of 370 days, and the color indexes of the Gaia DR2/EDR3 sources. The Gaia DR2 source ID 4097002430790019712 is classified as a large amplitude variable with an amplitude $\Delta G$ = 0.06 mag. The Bochum Galactic Disk Survey II refers to this variable as source ID GDS J1825392-163345, with a median light curve magnitude r = 16.51, i = 14.22 and a maximum amplitude of 4.10 mag. The ASAS-SN Catalog of Variable Stars II classifies this star, ASASSN-V J182539.23-163346.0, as a Semiregular, with a mean magnitude V = 15.13, an amplitude of 0.59 mag and a period of 340 days.

Our period analysis, based on 681 and 791 valid observations available from ASAS-SN in the V and g filters, applying the Lomb-Scargle method highlighted three potential periods in the range from 122 to 379 days. Two solutions are close to the original value and their weighted average is 371 ± 10 days. In addition, we found a second potential period at 122 ± 3 days (see Figure 14).

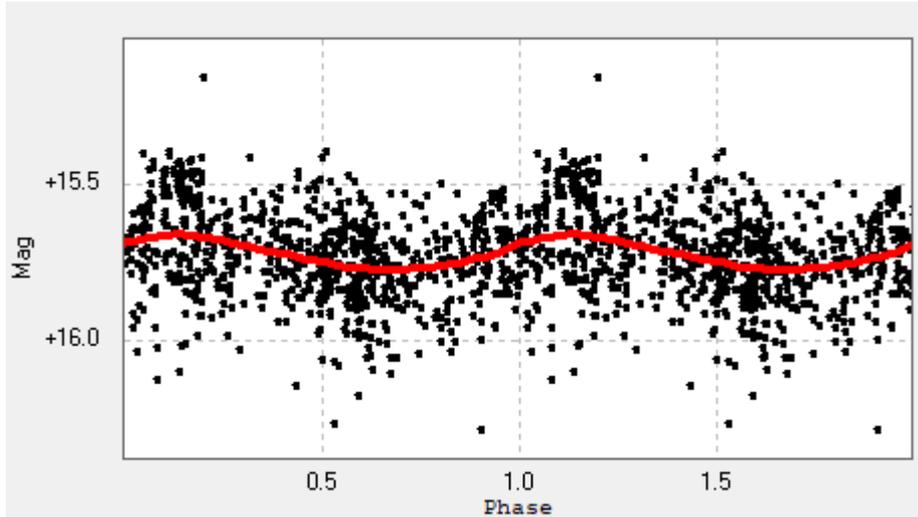

Figure 14     V3942 Sgr - V mag vs Phase (122 ± 3 days period - Lomb-Scargle)

**V3943 Sgr**

This star is identified with the infrared counterpart 2MASS J18264699-1557304 and Gaia DR2/EDR3 source ID 4097889461785848576. SIMBAD and AAVSO database report this star as a Mira in accordance with the original study, that determined a period of 374 days. This classification is consistent with the Gaia DR2 color index Bp-Rp = 6.502 and absolute magnitude $M_G$ = 2.93 (-1.36; +1.19) which collocate this variable in the LPV group of Figure 1.

It is classified as a large amplitude variable with an amplitude ΔG = 0.74 mag and is reported as a WISE J182646.99-155730.3 candidate YSO.

The Bochum Galactic Disk Survey II refers to this variable as source ID GDS J1826471-155730, with a median light curve magnitude i = 15.85 and a maximum amplitude of 3.66 mag. The ASAS-SN Catalog of Variable Stars II does not report any light curve for this variable. We did not perform any analysis of the ASAS-SN photometric data due to insufficient valid data available.

**V3944 Sgr**

This star is identified with the infrared counterpart 2MASS J18264776-1621427 and Gaia DR2/EDR3 source ID 4097087883482681344. SIMBAD and AAVSO database report this star as a Mira in accordance with the original study, that determined a period of 286 days.

This classification is consistent with the Gaia DR2 color index Bp-Rp = 5.591 and absolute magnitude $M_G$ = 2.01 ± 0.99, which collocate this variable in the LPV group of Figure 1.

It is classified as a large amplitude variable with an amplitude ΔG = 0.61 and is reported as a WISE J182647.78-162142.4 candidate YSO.

The Bochum Galactic Disk Survey II refers to this variable as source ID GDS J1826478-162142, with a median light curve magnitude i = 15.84 and a maximum amplitude of 2.87 mag. The ASAS-SN Catalog of Variable Stars II classifies this star, ASASSN-V J182647.77-162142.7, as a Semiregular, with a mean magnitude V = 16.49, an amplitude of 0.43 mag and a period of 19 days. We did not perform any analysis of the ASAS-SN photometric data due to insufficient valid data available.

**V3945 Sgr**

The variable star V3945 Sgr is classified as a Mira candidate and identified with the infrared counterpart 2MASS J18272269-1615389 and source ID Gaia DR2/EDR3 4097096056729738624. The Gaia DR2 color index Bp-Rp = 5.737 and absolute magnitude $M_G$ = 0.57 (-0.91; +0.93) are consistent with a LPV, as shown in Figure 1.

It is classified as a large amplitude variable with an amplitude $\Delta G$ = 0.94 and the Bochum Galactic Disk Survey II refers to this variable as source ID GDS J1827227-161538, with a median light curve magnitude i = 15.56 and a maximum amplitude of 1.95 mag.

The ASAS-SN Catalog of Variable Stars II identifies this star as ASASSN-V J182722.70-161538.9, and classifies it as Semiregular, with a mean magnitude V = 16.58, an amplitude of 0.81 mag and a period of 320 days, in accordance with the original value.

We performed a period analysis of the ASAS-SN photometric data but due to the insufficient valid photometric data available we did not find any reliable period.

**V3946 Sgr**

This star is identified with the infrared counterpart 2MASS J18274130-1621546 and the Gaia DR2/EDR3 source ID 4097090769700294656. SIMBAD and AAVSO database report this variable as a Mira in accordance with the original study, which determined a period of 411 days. This classification is consistent with the Gaia DR2 color index Bp-Rp = 5.747 and absolute magnitude $M_G$ = 2.25 ± 1.16, which collocate this variable in the LPV group of Figure 1.

It is classified as a large amplitude variable with an amplitude $\Delta G$ = 0.77 and reported as a WISE J182741.27-162154.4 candidate YSO.

This variable shows also a short-time sudden increase of magnitude, which occurs at phase = 0.96, with an amplitude of 0.7 mag (I-N hypersensitized+RG5) and a duration of 3.0 days (Maffei and Tosti, 1995).

The ASAS-SN Catalog of Variable Stars II does not report any light curve for this variable.

Our period analysis, based on 580 valid observations available from ASAS-SN in the V filter, applying the ANOVA method did not highlight reliable periods.

**V3947 Sgr**

This star is identified with the infrared counterpart 2MASS J18274739-1607056 and the Gaia DR2/EDR3 source ID 4097121070619720704. SIMBAD and AAVSO database report this variable as a Mira in accordance with the original study, which determined a period of 314.5 days. This classification is consistent with the Gaia DR2 color index Bp-Rp = 6.090 and absolute magnitude $M_G$ = 0.12 (-1.01; +1.03), which collocate this variable in the LPV group of Figure 1. The Gaia DR2 classifies this object as a candidate LPV, with a period of 285 ± 26 days and an amplitude of 0.64 mag. The Bochum Galactic Disk Survey II refers to this variable as source ID GDS J1827474-160705, with a median light curve magnitude r = 16.13, i = 13.59 and a maximum amplitude of 3.76 mag.

The ASAS-SN Catalog of Variable Stars II identifies this star, ASASSN-V J182747.62-160709.6, with a different Gaia EDR3 source ID 4097121074968426752 and classifies it as YSO, with a mean magnitude V = 13.97 and an amplitude of 0.34 mag. The ASAS-SN cross-reference ID is deemed incorrect because source ID 4097121074968426752 is a fainter star (G = 18.292) 5 arcsec from 4097121070619720704.

Our period analysis, based on valid observations available from ASAS-SN in the V and g filters, applying the Lomb-Scargle and ANOVA methods highlighted several potential solutions around 30 days. The weight average of the values we found is 29.5 ± 0.1 (see Figure 15). No reliable periods were found around the original and LPV results of 300 days.

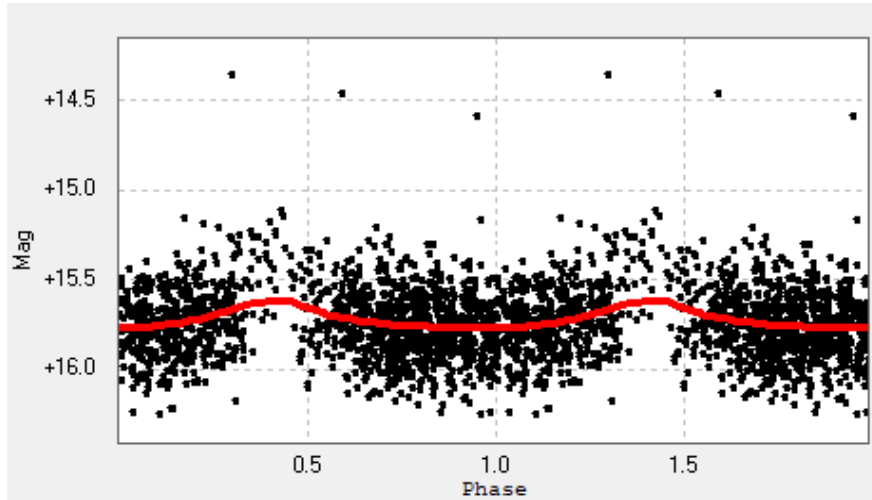

Figure 15    V3947 Sgr - V mag vs Phase (29.5 ± 0.1 days period – weighted average)

**V3948 Sgr**

This star is identified with the infrared counterpart 2MASS J18275269-1622048 and Gaia DR2/EDR3 source ID 4097090632223485824. SIMBAD and AAVSO database report this variable as a Mira in accordance with the original study, which determined a period of 420 days. This classification is consistent with the Gaia DR2 color index Bp-Rp = 5.036 and absolute magnitude $M_G$ = 2.73 (-1.16; +1.08), which collocate this variable in the LPV group of Figure 1. It is classified as a large amplitude variable with an amplitude $\Delta G$ = 1.05 and reported as a WISE J182752.68-162204.8 candidate YSO.

The ASAS-SN Catalog of Variable Stars II identifies this star as ASASSN-V J182752.70-162204.8 and classifies it as Semiregular, with a mean magnitude V = 15.34, an amplitude of 0.17 mag and a period of 18 days.

Our period analysis, based on 472 valid observations available from ASAS-SN in the V filter, applying the Lomb-Scargle method did not highlight any reliable period.

**V3949 Sgr**

This star is identified with the infrared counterpart 2MASS J18281973-1621109 and Gaia DR2/EDR3 source ID 4097067336305154304. SIMBAD and AAVSO database report this variable as a Mira in accordance with the original study, which determined a period of 520 days. This classification is consistent with the Gaia DR2 color index Bp-Rp = 6.614 and absolute magnitude $M_G$ = 0.32 (-1.04; +1.08), which collocate this variable in the LPV group of Figure 1.

This variable is classified as a LPV candidate by Gaia DR2 with a period of 485 ± 154 days and an amplitude of 0.44 mag. The Bochum Galactic Disk Survey II refers to this variable as source ID GDS J1828197-162111, with a median light curve magnitude r = 16.88, i = 14.77 and a maximum amplitude of 3.35 mag. The ASAS-SN Catalog of Variable Stars II identifies this star as ASASSN-V J182819.74-162111.0 and classifies it as a Semiregular, with a mean magnitude V = 15.95, an amplitude of 0.44 mag and a period of 449 days.

Our period analysis, based on 645 valid observations available from ASAS-SN in the V filter, applying the Lomb-Scargle and ANOVA methods did not highlight any reliable period.

**V3950 Sgr**

This star is identified the infrared counterpart 2MASS J18283838-1603253 (Nesci, 2018) and Gaia DR2/EDR3 source id 4097128324897371264. SIMBAD and AAVSO database report this variable as a LPV candidate and a Semiregular respectively, in accordance with the original study, which determined an uncertain period of 198 days. This classification is consistent with the Gaia DR2 color index Bp-Rp = 4.887 and the absolute magnitude $M_G$ = 1.14 (-1.03; +1.00), which place this star in the LPV group of Figure 1. It is noted that SIMBAD database incorrectly refers to source 2MASS J18283821-1603286 that is located 4" southwest of V3950 Sgr.

It is classified a large amplitude variable with an amplitude of 0.15 mag and the Bochum Galactic Disk Survey II refers to this variable as source ID GDS J1828384-160325, with a median light curve magnitude r = 16.06, i = 13.95 and a maximum amplitude of 1.34 mag.

The ASAS-SN Catalog of Variable Stars II does not report any light curve for this variable.

We performed our period analysis on 595 valid observations available from ASAS-SN in the V filter, on a time span of 1313 days, applying both Lomb-Scargle and ANOVA methods. The weighted average of the two periods we obtained with the two methods is 197 ± 10 days, in accordance with the original study (see Figure 16).

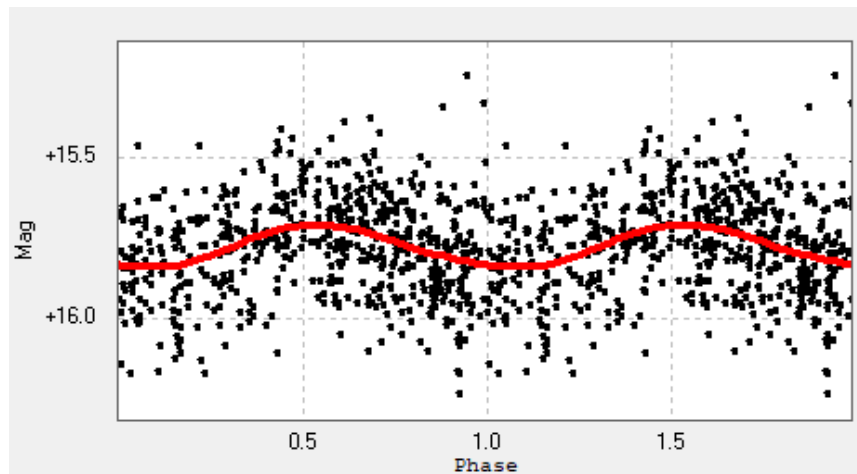

Figure 16   V3950 Sgr - V mag vs. Phase (197 ± 10 days period – weighted average)

## 3. Conclusions

The assessment of data available from public astronomical databases and referenced papers, and the analysis we performed on light curves and periods, refine the original classification, or suggest a revision of the type and/or the period for several of the 40 variables, part of the Sagittarius constellation of IBVS 985.

For all stars, the variability is confirmed.

For the following 15 stars, our analysis provides refinements or solutions for the type and/or period significantly different from existing studies: NSV 10490, NSV 10522, NSV 10741, NSV 10772, V3904 Sgr, V3920 Sgr, V3921 Sgr, V3927 Sgr, V3930 Sgr, V3932 Sgr, V3934 Sgr, V3937 Sgr, V3939 Sgr, V3942 Sgr, V3947 Sgr.

Our assessment also identifies 9 cases for which characteristics available from ASAS-SN photometric database are significantly different from the original classification. The variables NSV 10577, V3904 Sgr, V3924 Sgr, V3930 Sgr, V3935 Sgr, V3936 Sgr, V3937 Sgr, V3944 Sgr, V3948 Sgr are all classified as Mira in the original study, with periods in the range from 286 to 665 days, whilst ASAS-SN database classifies them as Semiregular with shorter periods in the range from 6 to 57 days.

We noted this discrepancy in a previous work (La Rocca et al. 2021) where original light curves were not available. Due to the good quality of the available original Mira light curves we deem that this discrepancy is due to the rules applied by the ASAS-SN system for the classification of the types and the period determination.

Our assessment also highlights that for 5 stars incorrect cross-reference names are reported by ASAS-SN Catalog of Variable Stars and a misidentification of NSV 10522 of the 2MASS infrared counterpart is reported in IBVS 6255

Table 7 provides a summary of the main results of this work.

Table 7        Summary of the main results of this work

| (Maffei and Tosti, 2013) | | | This assessment | | |
|---|---|---|---|---|---|
| **Variable** | **Type** | **Period (d)** | **Type** | **Period (d)** | **Remarks** |
| NSV 10490 | E:: | --- | Probable YSO (our analysis) | Not applicable (our analysis) | --- |
| | | | SR: (AAVSO) | --- (AAVSO) | |
| NSV 10522 | --- | --- | Probable YSO (our analysis) | 7.475 ± 0.012 (our analysis) | Maximum = 2457842 ± 1 HJD (our analysis)<br><br>Incorrect cross reference to Gaia EDR3 4097214842648260096 (our analysis) |
| | | | YSO (ASAS-SN) | Not applicable (ASAS-SN) | Misidentification of 2MASS infrared counterpart in IBVS 6255. Correct ID for NSV 10522 is 2MASS 18182915-1725379 (our analysis) |

| (Maffei and Tosti, 2013) | | | This assessment | | |
|---|---|---|---|---|---|
| **Variable** | **Type** | **Period (d)** | **Type** | **Period (d)** | **Remarks** |
| NSV 10577 (V5514 Sgr) | M | 665.0 | Not classified (our analysis) | --- (our analysis) | Gaia EDR3 photometric characteristics are consistent with a LPV star (our analysis) |
| | | | SR (ASAS-SN) | 15 (ASAS-SN) | Values of median magnitude G and color index Bp-Rp in Gaia DR2 and EDR3 > 1.3 (our analysis) |
| NSV 10626 (V5532 Sgr) | M: | 338.0:: | Not classified (our analysis) | --- (our analysis) | Gaia EDR3 photometric characteristics are consistent with a LPV star (our analysis) |
| NSV 10693 | --- | --- | LPV (our analysis) | 18 ÷ 423 (our analysis) | Maximum = 2457936 ± 4 HJD (our analysis) |
| | | | SR (ASAS-SN) | 701 (ASAS-SN) | |
| NSV 10741 | --- | --- | SR (our analysis) | 227 ± 7 (our analysis) | Maximum = 2457881 ± 7 HJD (our analysis) <br><br> Incorrect cross reference to Gaia EDR3 4096948550378843264 (our analysis) |
| | | | VAR (ASAS-SN) | 230 (ASAS-SN) | Redder and fainter red giant (2MASS J18242495-1703556, Gaia DR2/EDR3 4096948550378843264; G= 16.0) 8" SW (AAVSO) and classified as a large amplitude variable (ΔG = 0.65 mag, Gaia DR2). |
| NSV 10752 (V5536 Sgr) | M | 206.0 | LPV (our analysis) | --- (our analysis) | --- |
| | | | Candidate YSO (WISE) | Not applicable (WISE) | |
| NSV 10757 | --- | --- | Not classified (our analysis) | --- (our analysis) | Gaia DR2 4097858602406503552 (G = 15.755) and Gaia EDR3 4097858606748077696 (G = 17.134) (our analysis) |
| | | | Candidate YSO (WISE) | Not applicable (WISE) | |
| NSV 10772 | --- | --- | LPV (our analysis) | 15 ÷ 700 (our analysis) | Maximum = 2459073 ± 3 HJD (our analysis) |
| | | | Candidate YSO (WISE) | Not applicable (WISE) | |
| | | | SR: (AAVSO) | --- (AAVSO) | |
| | | | L (ASAS-SN) | --- (ASAS-SN) | |

| (Maffei and Tosti, 2013) | | | This assessment | | |
|---|---|---|---|---|---|
| **Variable** | **Type** | **Period (d)** | **Type** | **Period (d)** | **Remarks** |
| NSV 10837 (V5538 Sgr) | SRa | 416.0 | Not classified (our analysis) <br><br> Candidate YSO (WISE) <br><br> VAR (ASAS-SN) | Not reliable period found (our analysis) <br><br> Not applicable (WISE) <br><br> --- (ASAS-SN) | --- |
| V3904 Sgr | M | 360.0 | LPV (our analysis) <br><br> SR (ASAS-SN) | 29.6 ± 0.4 (our analysis) <br><br> 57 (ASAS-SN) | --- |
| V3905 Sgr | SRa | 328.0 | LPV (our analysis) <br><br> Candidate YSO WISE | --- (our analysis) <br><br> Not applicable (WISE) | --- |
| V3908 Sgr | M | 274.5 | LPV (our analysis) <br><br> M (ASAS-SN) <br><br> LPV Candidate (Gaia DR2) | 281 ± 1 (our analysis) <br><br> 281 (ASAS-SN) <br><br> 269 ± 17 (Gaia DR2) | Maximum = 2457863 ± 1 HJD (our analysis) |
| V3918 Sgr | M | 384.0 | LPV (our analysis) | --- (our analysis) | --- |
| V3920 Sgr | SR:: | --- | Probable YSO (our analysis) <br><br> VAR (ASAS-SN) | Not applicable (our analysis) <br><br> --- (ASAS-SN) | Incorrect cross reference to Gaia EDR3 4097209551260717312 (our analysis) |
| V3921 Sgr | M | 531.0 | Probable YSO (our analysis) | Not applicable (our analysis) | --- |
| V3923 Sgr | M:: | --- | LPV (our analysis) <br><br> SR (ASAS-SN) | --- (our analysis) <br><br> 8 (ASAS-SN) | --- |
| V3924 Sgr | M | 372.0 | LPV (our analysis) <br><br> Candidate YSO (WISE) <br><br> SR (ASAS-SN) | Not reliable period found (our analysis) <br><br> Not applicable (WISE) <br><br> 6 (ASAS-SN) | --- |

| (Maffei and Tosti, 2013) | | | This assessment | | |
|---|---|---|---|---|---|
| **Variable** | **Type** | **Period (d)** | **Type** | **Period (d)** | **Remarks** |
| V3925 Sgr | M | 151.5 | LPV (our analysis) | --- (our analysis) | --- |
| | | | Candidate YSO (WISE) | Not applicable (WISE) | |
| | | | LPV Candidate (Gaia DR2) | 148 ± 8 (Gaia DR2) | |
| V3926 Sgr | M | 290.0 | LPV (our analysis) | Not reliable period found (our analysis) | --- |
| | | | SR (ASAS-SN) | 274 (ASAS-SN) | |
| V3927 Sgr | SR | 218.0 | LPV (our analysis) | 7 ÷ 219 (our analysis) | --- |
| | | | SR (ASAS-SN) | 257 (ASAS-SN) | |
| V3930 Sgr | M | 448.0 | Probable YSO (our analysis) | Not applicable (our analysis) | --- |
| | | | Candidate YSO (WISE) | Not applicable (WISE) | |
| | | | SR (ASAS-SN) | 7 (ASAS-SN) | |
| V3931 Sgr | M | 441.0 | LPV (our analysis) | --- (our analysis) | --- |
| V3932 Sgr | M | 205.0 | LPV (our analysis) | 58.4 ± 0.3 (our analysis) | --- |
| | | | SR (ASAS-SN) | 201 (ASAS-SN) | |
| V3934 Sgr | SRa | 366.0 | Not LPV (our analysis) | 29 ÷ 188 (our analysis) | Incorrect cross reference to Gaia EDR3 4096561659747632000 (our analysis) |
| | | | VAR (ASAS-SN) | 181 (ASAS-SN) | |
| V3935 Sgr | M | 395.0 | LPV (our analysis) | Not reliable period found (our analysis) | --- |
| | | | Candidate YSO (WISE) | Not applicable (WISE) | |
| | | | SR (ASAS-SN) | 15 (ASAS-SN) | |
| V3936 Sgr | M | 369.0 | LPV (our analysis) | 319 ± 47 (our analysis) | --- |
| | | | SR (ASAS-SN) | 44 (ASAS-SN) | |

| (Maffei and Tosti, 2013) | | | This assessment | | |
|---|---|---|---|---|---|
| **Variable** | **Type** | **Period (d)** | **Type** | **Period (d)** | **Remarks** |
| V3937 Sgr | M | 606.0 | Probable YSO (our analysis)<br><br>SR (ASAS-SN) | Not applicable (our analysis)<br><br>13 (ASAS-SN) | --- |
| V3938 Sgr | M | 297.0 | LPV (our analysis)<br><br>Candidate YSO (WISE) | --- (our analysis)<br><br>Not applicable (WISE) | --- |
| V3939 Sgr | M: | 292.0 | Probable YSO (our analysis) | Not applicable (our analysis) | --- |
| V3940 Sgr | M | 449.0 | LPV (our analysis)<br><br>SR (ASAS-SN) | $444 \pm 24$ (our analysis)<br><br>449 (ASAS-SN) | Maximum = $2458961 \pm 6$ HJD (our analysis) |
| V3942 Sgr | M | 370.0 | Not classified (our analysis)<br><br>SR (ASAS-SN) | $122 \pm 3$<br>$371 \pm 10$ (our analysis)<br><br>340 (ASAS-SN) | --- |
| V3943 Sgr | M | 374.0 | LPV (our analysis)<br><br>Candidate YSO (WISE) | --- (our analysis)<br><br>Not applicable (WISE) | --- |
| V3944 Sgr | M | 286.0 | LPV (our analysis)<br><br>Candidate YSO (WISE)<br><br>SR (ASAS-SN) | --- (our analysis)<br><br>Not applicable (WISE)<br><br>19 (ASAS-SN) | --- |
| V3945 Sgr | M | 321.0 | LPV (our analysis)<br><br>SR (ASAS-SN) | Not reliable period found (our analysis)<br><br>320 (ASAS-SN) | --- |
| V3946 Sgr | M | 411.0 | LPV (our analysis)<br><br>Candidate YSO (WISE) | Not reliable period found (our analysis)<br><br>Not applicable (WISE) | --- |

| (Maffei and Tosti, 2013) | | | This assessment | | |
|---|---|---|---|---|---|
| **Variable** | **Type** | **Period (d)** | **Type** | **Period (d)** | **Remarks** |
| V3947 Sgr | M | 314.5 | LPV (our analysis) | 29.5 ± 0.1 (our analysis) | Incorrect cross reference to Gaia EDR3 4097121074968426752 (our analysis) |
| | | | Candidate LPV (Gaia DR2) | 285 ± 26 (Gaia DR2) | |
| | | | YSO (ASAS-SN) | Not applicable (ASAS-SN) | |
| V3948 Sgr | M | 420.0 | LPV (our analysis) | Not reliable period found (our analysis) | --- |
| | | | Candidate YSO (WISE) | Not applicable (WISE) | |
| | | | SR (ASAS-SN) | 18 (ASAS-SN) | |
| V3949 Sgr | M: | 520.0 | LPV (our analysis) | Not reliable period found (our analysis) | --- |
| | | | Candidate LPV (Gaia DR2) | 485 ± 154 (Gaia DR2) | |
| | | | SR (ASAS-SN) | 449 (ASAS-SN) | |
| V3950 Sgr | SRa | 198.0: | LPV (our analysis) | 197 ± 10 (our analysis) | --- |
| | | | Candidate YSO (WISE) | Not applicable (WISE) | |

**Acknowledgements**

- This activity has made use of the SIMBAD database, operated at CDS, Strasbourg, France.
- This work has made use of data from the European Space Agency (ESA) mission Gaia (https://www.cosmos.esa.int/gaia), processed by the Gaia Data Processing and Analysis Consortium (DPAC, https://www.cosmos.esa.int/web/gaia/dpac/consortium). Funding for the DPAC has been provided by national institutions, in particular the institutions participating in the Gaia Multilateral Agreement.
- We acknowledge with thanks the variable star observations from the *AAVSO International Database* contributed by observers worldwide and used in this research.
- This work was carried out in the context of educational and training activities provided by Italian law 'Percorsi per le Competenze Trasversali e l'Orientamento', December 30th, 2018, n.145, Art.1.